\documentclass[twocolumn,showpacs,amsmath,amssymb]{revtex4}
\usepackage{graphicx}
\usepackage{slashed}
\usepackage{pifont}
\usepackage{color}
\usepackage{epsfig}

\newcommand{\beqn}{\begin{equation}}
\newcommand{\eeqn}{\end{equation}}

\newcommand{\Real}{\mathtt{Re}}

\begin{document}
\title{Effects of Radiation-Reaction in Relativistic Laser Acceleration}
\author{Y. Hadad, L. Labun, J. Rafelski}
\affiliation{Departments of Physics and Mathematics, University of Arizona, Tucson, Arizona, 85721 USA}

\author{N. Elkina, C. Klier, H. Ruhl}
\affiliation{Department~f\"ur~Physik~der~Ludwig-Maximillians-Universit\"at,~Theresienstrasse~37A,~80333 M\"unchen,~Germany}

\date{14 November, 2010} 

\begin{abstract}
The goal of this paper is twofold: to explore the response of classical charges to electromagnetic force
at the level of unity in natural units and to establish a criterion that determines physical parameters for which
the related radiation-reaction effects are detectable. In pursuit of this goal, the Landau-Lifshitz equation is solved
analytically for an arbitrary (transverse) electromagnetic pulse. A comparative study of the radiation emission
of an electron in a linearly polarized pulse for the Landau-Lifshitz equation and for the Lorentz force equation
reveals the radiation-reaction-dominated regime, in which radiation-reaction effects overcome the influence
of the external fields. The case of a relativistic electron that is slowed down by a counterpropagating electromagnetic
wave is studied in detail. We further show that when the electron experiences acceleration of order unity,
the dynamics of the Lorentz force equation, the Landau-Lifshitz equation and the Lorentz-Abraham-Dirac equation
all result in different radiation emission that could be distinguished in experiment.  Finally, our analytic and numerical results are compared with those appearing in the literature.
\end{abstract}


\maketitle

\section{Introduction}
\subsection{The radiation-reaction force and Landau-Lifshitz equation}
The equation of motion for an electron with charge $-e$ and mass $m$ 
in an external electromagnetic field is given by the {\it Lorentz force ({\bf LF}) equation}
\beqn \label{LorentzEq}
m \dot{u}^\alpha=-e F^{\alpha\beta} u_\beta,
\eeqn
where $F^{\alpha\beta}$ is the electromagnetic tensor, $u^\alpha=\gamma(1,\vec{v})$ 
is the four-velocity of the charge, $\gamma={1}/{\sqrt{1-v^2}}$ and the dot represents 
differentiation with respect to proper time $\tau$. We use the metric convention 
$\left(+---\right)$ and units in which the speed of light is unity  $c=1$. Maxwell equations together
with the LF equation, imply that the rate at which energy is emitted by an accelerated charge
with respect to laboratory time is~\cite{Jackson}
\beqn \label{Radiation}
\mathcal{R} \left.\equiv \frac{dE}{dt}\right\vert_{\rm lab} = -\frac{2}{3} e^2 \dot{u}^\alpha \dot{u}_\alpha.
\eeqn
The right-hand-side of Eq.\,(\ref{Radiation}) is a positive-definite Lorentz invariant that vanishes if and only if $\dot{u}^\beta=0$,
and therefore an accelerated charge emits radiation and loses energy relative to any Lorentz observer. 

The inertial reaction due to the energy-momentum loss exhibited by Eq.\,\eqref{Radiation} is not accounted for in the LF equation~\cite{Dirac:1938nz}.
Consequently, the Lorentz force equation with a prescribed electromagnetic field is
only an approximated description of the electron motion limited to cases in which the radiation
emission is small. 
The precise meaning of ``small'' will be established in Sec.~\ref{RadiationSection}
where the {\it radiation-reaction ({\bf RR}) dominated regime criterion} suggested 
by analysis~\cite{LandauLifshitz,Koga2} is verified by numerical evaluation of the radiation.

During the development of the Lorentz-Maxwell theory of electromagnetism,
there has been a long search for an improved classical equation that 
comprehensively describes the motion of a radiating charge in a prescribed electromagnetic field.
The most prominent equation that was suggested after the introduction of quantum theory
is the Lorentz-Abraham-Dirac ({\bf LAD}) equation~\cite{Dirac:1938nz}
\beqn \label{LADEq}
m\dot{u}^\alpha=-eF^{\alpha\beta}u_{\beta}+m\tau_0 \left[\ddot{u}^\alpha+\dot{u}^2 u^\alpha\right],
\eeqn
where
\beqn \label{Tau0}
\tau_0=\frac{2}{3}\frac{e^2}{mc^3}
\eeqn
is a constant with dimensions of time, whose numerical value for the electron 
is $\tau_0=6.24\times 10^{-24}\,{\rm s}$.  Other models have been introduced by 
Eliezer~\cite{Eliezer}, Landau and Lifshitz~\cite{LandauLifshitz}, Mo and Papas~\cite{TseChin:1972kg}, 
Caldirola~\cite{Caldirola:1979mi}, Yaghjian~\cite{Yaghjian} and Sokolov et al.~\cite{Sokolov:2009rl,Sokolov:2009ee,Sokolov:2010jx}.

Of particular interest is the {\it Landau-Lifshitz ({\bf LL}) equation} (also known as the
reduced LAD equation) first presented in~\cite{LandauLifshitz}
\begin{eqnarray} \label{LLEq}
m\dot{u}^{\alpha} \! &=& \! -e F^{\alpha\beta}u_{\beta}-e\tau_{0}\bigg\{F_{,\gamma}^{\alpha\beta}u_{\beta}u^{\gamma}\notag \\
&&-\frac{e}{m}\left[F^{\alpha\beta} F_{\beta\gamma} u^\gamma - F^{\beta\gamma} F_{\gamma\delta}u^\delta u_{\beta} u^\alpha\right]\bigg\}.
\end{eqnarray}
The LL equation is often selected for further study because it is the only equation from the above list that is equivalent to the LAD equation up to first order in $\tau_0$ \cite{Rohrlich2}, avoids the nonphysical solutions of the LAD ~\cite{Spohn:1999uf} and has known analytic solutions~\cite{Rivera,Rajeev:2008sw,DiPiazza}.  A rigorous derivation of Eq.\,\eqref{LLEq} using perturbation theory was recently given from considerations of energy and momentum conservation~\cite{Gralla:2009md}.  

\subsection{Line of approach and objectives}
In this paper we explore the dynamics of a charged particle exposed to an
ultraintense pulsed laser field. The objective will be to 
understand how RR impacts the normal Lorentz dynamics and to derive criteria which 
delineate the domain of validity of the LF. For a thorough discussion of the electron dynamics
and radiation emission in laser fields with a small radiation reaction correction, the reader
should consult Ref.~\cite{Sarachik:1970ap}.  The assumption of small radiation-reaction breaks down as the acceleration reaches unity in natural units.
For acceleration unity, the familiar Lorentz force dynamics cannot hold anymore, and RR effects dominate
the motion of the charge as will be discussed in depth in the body of this work.

The investigations here are intended to help identify the onset of new dynamics beyond that found in classical electromagnetism, and not so much to test which of the LF generalizations is more accurate. Moreover, since the theory of
quantum electrodynamics is founded upon classical electromagnetism, the 
completeness and depth of our understanding of charged particles under high accelerations
remains both qualitatively and quantitatively uncertain~\cite{Rafelski:2009fi}.  In particular, dynamics at the critical 
electromagnetic field strength $E_c=m^2c^3/e\hbar$---both classical and quantum---should be reconsidered~\cite{Fedotov:2010ja}.  At this limit quantum effects are believed important~\cite{Ritus1985, Baier:1998vh}, though we withhold judgment on the matter 
and consider a thorough analysis of classical predictions a relevant basis for further exploration.

\begin{figure}[b!]
\begin{center}
\includegraphics[scale=0.4]{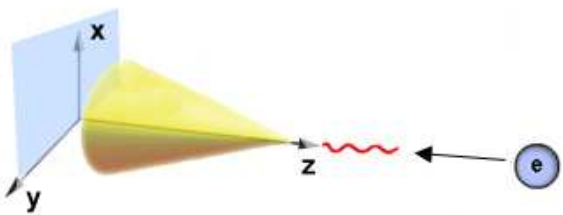}
\caption{\label{ExperimentScheme} The experimental setup, represented qualitatively.
The radiation propagates mostly in the direction of the electron and is projected onto the screen on the left.}
\end{center}
\end{figure}

To reach acceleration unity with current laser systems, we boost the intensity of the laser wave
by colliding it head-on with relativistic electrons (Fig. \ref{ExperimentScheme}). In the instantaneous rest
frame of the electron the laser fields are greatly enhanced, as was demonstrated numerically in \cite{Koga2,Koga1}.
Since the LL equation is considered to be the best available
candidate to account for the radiation-reaction effects, we solve the LL equation and analyze the dynamics according
to this model.

The LL Eq.\,(\ref{LLEq}) is nonlinear in the electromagnetic field 
tensor $F^{\alpha\beta}$ and in the four-velocity $u^\alpha$. Because of this nonlinearity, 
it has resisted for a long time an analytic solution, except for simple cases:
constant magnetic~\cite{Herrera:1973zz,Herrera:1974py} and constant magnetic plus electric fields for which the LL equation reproduces the dynamics of the LAD equation~\cite{deParga}, a circular orbit~\cite{Rivera} and the nonrelativistic 
motion in a Coulomb potential~\cite{Rajeev:2008sw}. Recently~\cite{DiPiazza}, a closed-form solution for LL was obtained for a plane wave and studied in applications~\cite{DiPiazza2}.

In this paper we solve for the motion of a charged particle in a 
{\it transverse wave}: a monochromatic  electromagnetic wave for which 
the wave fronts (surfaces of constant phase) are infinite parallel planes 
of arbitrary amplitude. This includes many useful physical scenarios as special 
cases, such as constant crossed electromagnetic fields, linearly polarized 
and circularly polarized plane waves, and an electromagnetic pulse in space 
and time with arbitrary shape. 
The solution here is obtained by a method independent of that seen in~\cite{DiPiazza},
and we provide additionally an analysis of the radiation emission and its angular distribution.  Analytic and numerical comparisons of our solution with those of~\cite{DiPiazza,DiPiazza2} are provided in Appendix~\ref{app:solncomp}.

In Sec.~\ref{TransverseWaveSection} we discuss the physical meaning 
and mathematical properties of the transverse wave. Section~\ref{SolutionSection} 
includes the derivation of the analytic solution of the LL equation for this case. 
In Sec.~\ref{RadiationSection} we compare the radiation emission with or 
without the RR force, and discuss in what conditions 
the LL equation can be probed experimentally. In Sec.~\ref{ExamplesSection} 
we study our two main examples of linearly polarized and circularly polarized 
plane waves and compare them to the known solution of the LF equation.

\section{The Transverse Wave} \label{TransverseWaveSection}
We consider the motion of a charged particle in a {\it transverse wave}: a monochromatic 
electromagnetic wave for which the wave fronts are infinite parallel planes of arbitrary 
amplitude. The four-potential of such an electromagnetic wave is
\beqn \label{WavePotential}
A^\alpha (x) = A_0 \Real\left[\varepsilon^\alpha f\left(\xi\right)\right],
\eeqn
where $A_0$ is the maximal amplitude of the wave, $\varepsilon^\alpha$ is the (complex) polarization four-vector, $k^\alpha=\left(\omega,\vec{k}\right)$ is the propagation four-vector and $f$ is an arbitrary (complex) function of $\xi=k^{\alpha}x_{\alpha}\equiv k \cdot x$ that represents the shape of the wave.

To keep the dependence on the intensity of the electromagnetic wave explicit, we introduce the {\it normalized four-potential}
\beqn \label{NormalizedWavePotential}
\hat{A}^\alpha (x) = \Real\left[\varepsilon^\alpha f\left(\xi\right)\right]
\eeqn
so that the four-potential is $A^\alpha=A_0 \hat{A}^\alpha$. This will prove useful in the analysis of the radiation emission in Sec.~\ref{RadiationSection}.

We take a polarization four-vector $\varepsilon^\alpha$ and a propagation four-vector $k^\alpha$ that satisfy
\begin{eqnarray} \label{WaveRelations}
k^2 &=& 0, \\ \notag
|\varepsilon|^2 &=& -1,
\end{eqnarray}
and the {\it transverse condition}
\beqn \label{TransverseCondition}
k \cdot \varepsilon = 0.
\eeqn
This means that the polarization of the wave is orthogonal to the direction of wave propagation. Namely, the electric and magnetic fields are perpendicular to the direction of energy transfer.

The transverse condition Eq.\,(\ref{TransverseCondition}) guarantees that the electromagnetic potential satisfies the Lorenz gauge condition
\beqn \label{LorenzGaugeCondition}
\partial_\alpha A^\alpha = 0.
\eeqn
The electromagnetic field tensor is
\begin{eqnarray} \label{PlaneWaveField}
F^{\alpha\beta}(x)&=&\partial^\alpha A^\beta - \partial^\beta A^\alpha \\ \notag
&=& A_0 \Real\left[\left(k^\alpha \varepsilon^\beta - k^\beta \varepsilon^\alpha \right)f' \left(\xi \right)\right] \\ \notag
&=& k^\alpha A'^\beta - k^\beta A'^\alpha,
\end{eqnarray}
where the prime denotes differentiation with respect to the variable $\xi$. Equation (\ref{WaveRelations}) and the transverse condition Eq.\,(\ref{TransverseCondition}) imply that the field tensor satisfies the following identities
\begin{subequations} \label{FieldIdentities}
\begin{eqnarray} \label{FieldIdentity1}
k_\alpha F^{\alpha\beta} &=& 0, \\ 
\label{FieldIdentity2}
k_\alpha F^{\alpha\beta}_{,\gamma} & = & 0, \\ 
\varepsilon_\alpha F^{\alpha\beta} & = & -(\varepsilon  \cdot A') k^\beta, \\ 
\varepsilon_\alpha F^{\alpha\beta}_{,\gamma} & = & -(\varepsilon \cdot A'') k^\beta k_\gamma, \\ 
F^{\alpha\beta} u_{\beta} & = & (u\cdot A')k^\alpha -(k \cdot u) A'^\alpha, \\ 
F^{\alpha\beta} F_{\beta\gamma} & = & -k^\alpha k_\gamma (A')^2.
\end{eqnarray}
\end{subequations}
Equation\:(\ref{FieldIdentity1}) reveals that the potential (\ref{WavePotential}) generates a very special field configuration, in which the wave four-vector $k^\alpha$ is orthogonal to each of the row/column four-vectors of the Faraday field tensor $F^{\alpha\beta}$. We will see later that this implies that the four-acceleration $\dot{u}^\alpha$ is orthogonal to $k^\alpha$ in the case of the LF Eq.\,(\ref{LorentzEq}). However, the orthogonality of the four-acceleration and the wave four-vector will cease to hold once RR terms are included, in the case of the LL Eq.\,(\ref{LLEq}).

Two concrete examples of the transverse wave Eq.\,(\ref{WavePotential}) we will address in Sec.~\ref{ExamplesSection} are

a) {\bf Linearly polarized plane wave} propagating in the positive $z$ direction with polarization in the $x$ direction. In this case
\begin{eqnarray} \label{LPPlaneWave}
\varepsilon^\alpha &=& (0, 1, 0, 0), \\ \notag
k^\alpha &=& (\omega, 0, 0, k), \\ \notag
f(\xi) &=& \sin(\xi-\xi_0), \\ \notag
\vec{A} &=& -A_0 \sin(k z - \omega t+\xi_0) \hat{x}, \\ \notag
\vec{E} &=& -\omega A_0 \cos(k z - \omega t+\xi_0) \hat{x}, \\ \notag
\vec{B} &=& -k A_0 \cos(k z - \omega t+\xi_0) \hat{y},
\end{eqnarray}
where $\xi_0$ is the phase of the wave. The choice of phase has an important physical significance, as it determines the intensity and the direction of the electromagnetic wave as it initially hits the particle. 

b) {\bf Circularly polarized plane wave} propagating in the positive $z$ direction with positive helicity. In this case
\begin{eqnarray} \label{CPPlaneWave}
\varepsilon^\alpha &=& \frac{1}{\sqrt{2}}(0, 1, -i, 0), \\ \notag
k^\alpha &=& (\omega, 0, 0, k), \\ \notag
f(\xi) &=& \sqrt{2} e^{i (\xi-\xi_0)}, \\ \notag
\vec{A}&=&A_0 \left[\cos(kz-\omega t + \xi_0)\hat{x}-\sin(kz-\omega t + \xi_0)\hat{y}\right], \\ \notag
\vec{E}&=&-\omega A_0 \left[\sin(kz- \omega t+\xi_0)\hat{x}\right. \\ \notag
&& \left.+\cos(k z - \omega t+\xi_0)\hat{y}\right], \\ \notag
\vec{B}&=&-k A_0 \left[-\cos(kz- \omega t+\xi_0)\hat{x}\right. \\ \notag
&& \left.+\sin(k z - \omega t+\xi_0)\hat{y}\right].
\end{eqnarray}
This example of a circularly polarized plane wave also demonstrates the importance of allowing $\varepsilon^\alpha$ and $f(\xi)$ to be complex.

\section{The Solution} \label{SolutionSection}
The gist of the method used to solve the LL equation is to introduce a change of variables similar to the one used in~\cite{Meyer} for the LF Eq.\,(\ref{LorentzEq}). We use the phase $\xi=k\cdot x$ instead of the proper time $\tau$ as the independent variable. This means that $\xi$ is related to the proper time $\tau$ by the relation
\beqn \label{XiTauRelation}
\frac{d\xi}{d\tau}=k\cdot u
\eeqn
since the wave vector $k^\alpha$ is fixed. Writing the LL Eq.\,(\ref{LLEq}) in terms of $\xi$, one obtains
\begin{eqnarray} \label{LLXi} \hspace*{-0.7cm}\notag
\left(k\cdot u\right) u'^{\alpha} \!\!&=&\!\!-\frac{e}{m} F^{\alpha\beta}u_{\beta}-\frac{e}{m} \tau_{0}\Big\{F_{,\gamma}^{\alpha\beta}u_{\beta}u^{\gamma}  \\ 
&&\!\! \left. -\frac{e}{m} \left[F^{\alpha\beta} F_{\beta\gamma} u^\gamma - u_{\beta}F^{\beta\gamma} F_{\gamma\delta}u^\delta u^\alpha\right]\right\}\!,
\end{eqnarray}
where the prime denotes differentiation with respect to the new variable $\xi$. Equation~\:(\ref{LLXi}) is a differential equation in $u^\alpha$ that contains the terms $\varepsilon \cdot u$ and $k \cdot u$ as we can see from identities (\ref{FieldIdentities}). Contracting Eq.\,(\ref{LLXi}) with $k_\alpha$ and using the field identities (\ref{FieldIdentities}) gives
\beqn \label{kuEqn}
(k\cdot u)' = \tau_0 a_0^2 \left(k \cdot u\right)^2 (\hat{A}')^2,
\eeqn
where 
\beqn \label{a_0}
a_0=\frac{eA_0}{m}
\eeqn
is a positive dimensionless constant measuring the intensity of the electromagnetic wave.

Dividing by $(k \cdot u)^2$ and integrating, we have
\beqn \label{ku}
k \cdot u = \frac{k \cdot u_0}{1-\tau_0 a_0^2 (k\cdot u_0) \psi(\xi)},
\eeqn
where we defined the first structure integral
\beqn \label{psiStructure}
\psi(\xi)=\int_0^\xi \left[\hat{A}'(y)\right]^2 dy.
\eeqn

Relation (\ref{XiTauRelation}) allows integration of Eq.\,(\ref{ku}), obtaining an explicit expression for $\tau$ as a function of $\xi$, namely
\beqn \label{XiTauLL}
\tau(\xi)=\frac{\xi}{k \cdot u_0}-\tau_0 a_0^2 \int_0^\xi \psi(y)dy.
\eeqn

If $A'^\alpha$ is a spacelike vector (e.g., when the time component of the polarization four-vector vanishes, $\varepsilon^0=0$), then when $\xi\geq 0$ the function $\psi$ is a non-negative function. Thanks to Eq.\,(\ref{ku}) we now see that $\frac{d\xi}{d\tau}>0$. Therefore $\xi$ is an (increasing) monotone function of $\tau$, and the change of variables $\tau\rightarrow\xi$ can indeed be used for $\tau\geq 0$ (notice, however, that it might run into a singularity if the proper time is negative).

There is a caveat in Eq.\,(\ref{ku}) that the reader should be aware of. Since we perform the integration with respect to $\xi$ and not $\tau$, the constant of integration should be determined by setting $\xi=0$. Given $u_0=u(\xi=0)$, in general $u_0$ is not the initial four-velocity of the particle. In order to remedy this situation and to minimize confusion, we now choose the coordinate system such that at $\tau=0$, the particle is at the origin and therefore $\xi=0$ as well. Since the change of variables was proved to be one-to-one, this guarantees that $u_0=u(\xi=0)=u(\tau=0)$, so for this particular coordinate system $u_0$ is indeed the initial four-velocity of the particle.

We continue by contracting Eq.\,(\ref{LLXi}) with $\varepsilon_\alpha$ and using the field identities (\ref{FieldIdentities}) in a similar fashion, which gives
\begin{eqnarray} \label{euEqn}
(k \cdot u)(\varepsilon \cdot u)' \!\!&=&\!\! a_0 (k \cdot u) (\varepsilon \cdot \hat{A}') \notag 
 + \tau_0 a_0 (k\cdot u)^2 (\varepsilon \cdot \hat{A}'') \\ 
&&+ \tau_0 a_0^2 (\varepsilon \cdot u)(k \cdot u)^2 (\hat{A}')^2.
\end{eqnarray}
This is a nonhomogeneous linear differential equation for $\varepsilon \cdot u$ which can be solved given $k\cdot u$. The last term in Eq.\,(\ref{euEqn}) is $(\varepsilon \cdot u)(k \cdot u)'$ as we can see from Eq.\,(\ref{kuEqn}). Therefore, one can write
\begin{eqnarray}
(k \cdot u)(\varepsilon \cdot u)'\!&-&\!(k \cdot u)'(\varepsilon \cdot u) = \\ \notag
&&  a_0 (k\cdot u) (\varepsilon \cdot \hat{A}') + \tau_0 a_0 (k\cdot u)^2 (\varepsilon \cdot \hat{A}''),
\end{eqnarray}
and dividing by $(k \cdot u)^2$ we have
\beqn
\left(\frac{\varepsilon \cdot u}{k \cdot u}\right)' = a_0 \frac{\varepsilon \cdot \hat{A}'}{k\cdot u}+\tau_0 a_0 \varepsilon \cdot \hat{A}''.
\eeqn
\\
Substituting Eq.\,(\ref{ku}) for $k \cdot u$ and integrating yields
\begin{eqnarray} \label{eu}
\frac{\varepsilon \cdot u}{k \cdot u} &=& \frac{\varepsilon \cdot u_0}{k \cdot u_0} + a_0 \frac{\varepsilon \cdot (\hat{A}-\hat{A}_0)}{k \cdot u_0}- \tau_0 a_0^3  \varepsilon \cdot \chi \\ \notag
&&+ \tau_0 a_0 \varepsilon \cdot (\hat{A}'-\hat{A}'_0),
\end{eqnarray}
where here again $\hat{A}^\alpha_0=\hat{A}^\alpha(0)$ and we defined as a second structure integral the four-vector
\beqn \label{chiStructure}
\chi^\alpha(\xi)= \int_0 ^\xi \hat{A}'^\alpha(y)\psi(y)dy.
\eeqn
Multiplying Eq.\,(\ref{eu}) by $f'$ and $f''$ and taking the real part gives us
\begin{eqnarray} \label{A'u}
\frac{\hat{A}' \cdot u}{k \cdot u} &=& \frac{\hat{A}'\cdot u_0}{k \cdot u_0} + a_0 \frac{\hat{A}' \cdot (\hat{A}-\hat{A}_0)}{k \cdot u_0} \\ \notag
&& -\tau_0 a_0 ^3 (\hat{A}' \cdot \chi)+\tau_0 a_0 \hat{A}' \cdot (\hat{A}'-\hat{A}'_0)
\end{eqnarray}
and
\begin{eqnarray} \label{A''u}
\frac{\hat{A}'' \cdot u}{k \cdot u} &=& \frac{\hat{A}'' \cdot u_0}{k \cdot u_0} + a_0\frac{\hat{A}'' \cdot (\hat{A}-\hat{A}_0)}{k \cdot u_0} \\ \notag
&& -\tau_0 a_0^3 (\hat{A}'' \cdot \chi) + \tau_0 a_0 \hat{A}'' \cdot (\hat{A}'-\hat{A}'_0).
\end{eqnarray}

We are finally ready to integrate the LL Eq.\,(\ref{LLXi}). Writing it explicitly in terms of the four-potential using Eq.\,(\ref{PlaneWaveField}) gives
\begin{eqnarray} \label{LLPlaneWave}
(k \cdot u) u'^{\alpha}&=&-a_0 \left[(\hat{A}' \cdot u) k^\alpha - (k \cdot u) \hat{A}'^\alpha \right] \\ \notag
&& -\tau_0 a_0 \left[(k \cdot u)(\hat{A}'' \cdot u) k^\alpha - (k\cdot u)^2 \hat{A}''^\alpha\right] \\ \notag
&& -\tau_0 a_0^2 \left[(k \cdot u)\hat{A}'^2 k^\alpha-(k\cdot u)^2 \hat{A}'^2 u^\alpha \right].
\end{eqnarray}
Similarly to what we had earlier, Eq.\,(\ref{kuEqn}) reveals that the last term is $(k\cdot u)' u^\alpha$, therefore
\begin{eqnarray} \label{LLPlaneWave2}
\hspace*{-0.8cm}\left(\!\frac{u^\alpha}{k\cdot u}\!\right)^{\!\!\prime}
\!\!&=&\!\!\!\left[a_0 \frac{1}{k\cdot u} \hat{A}'^\alpha + \tau_0 a_0 \hat{A}''^\alpha\right]  \notag \\ 
&& \!\!\!-\bigg[ a_0\frac{\hat{A}' \cdot u}{(k \cdot u)^2}+ \tau_0a_0 \frac{\hat{A}'' \cdot u}{k \cdot u} +\tau_0 a_0^2 \frac{\hat{A}'^2}{k \cdot u}\bigg] k^\alpha,
\end{eqnarray}
where we separated terms that are parallel to $\varepsilon^\alpha$ in the first brackets from terms that are parallel to $k^\alpha$ in the second brackets.
Substituting Eqs. (\ref{ku}), (\ref{A'u}) and (\ref{A''u}), integrating and collecting powers of $\tau_0$, we finally obtain
\begin{widetext}
\begin{eqnarray} \label{uLandauLifshitz}
u^\alpha&=&\frac{k\cdot u}{k\cdot u_0}\bigg\{u^\alpha_0+a_0 (\hat{A}^\alpha-\hat{A}^\alpha_0)-\frac{k^\alpha}{k \cdot u_0}\bigg[a_0 (\hat{A}-\hat{A}_0)\cdot u_0 +a_0 ^2 \frac{(\hat{A}-\hat{A}_0)^2}{2}\bigg]  \bigg\} + (k \cdot u) \tau_0 \bigg\{\left[a_0(\hat{A}'^\alpha-\hat{A}'^\alpha_0)-a_0 ^3 \chi^\alpha\right] \notag \\ 
&&\hspace*{0.8cm} +\frac{k^\alpha}{k\cdot u_0}\left[-a_0 (\hat{A}'-\hat{A}'_0) \cdot u_0 -a_0 ^2 \psi -a_0 ^2 (\hat{A}-\hat{A}_0)\cdot(\hat{A}'-\hat{A}'_0)+a_0^3 \chi \cdot u_0 +a_0^4 (\hat{A}-\hat{A}_0)\cdot \chi\right] \bigg\} \\ \notag
  &&  +k^\alpha(k \cdot u) \tau_0^2 \bigg\{-a_0^2 \frac{(\hat{A}'-\hat{A}'_0)^2}{2}+a_0^4 (\hat{A}'-\hat{A}'_0) \cdot \chi +a_0 ^4 \frac{\psi^2}{2} - a_0 ^6 \frac{\chi^2}{2}\bigg\}. 
\end{eqnarray}
\end{widetext}

This is an analytic expression for the four-velocity $u^\alpha$ as a function 
of the variable $\xi$. It is verified by direct computation that $u^\alpha u_\alpha=1$ 
as expected. Moreover, it is given in a manifestly covariant form that is valid 
for any reference frame. The presence of the four-potential $A^\alpha$ makes 
the solution not manifestly gauge invariant. Nevertheless, the solution is 
invariant under gauge transformations satisfying the transverse 
condition Eq.\,(\ref{TransverseCondition}), i.e. transformations 
of the form $A^\alpha(\xi) \rightarrow A^\alpha(\xi)+k^\alpha \Lambda(\xi)$.

Equation\:(\ref{uLandauLifshitz}) together with Eqs.\,(\ref{ku}) and (\ref{XiTauLL}) provide a complete description of the velocity of the particle as a function of its proper time. Once the structure integrals Eqs.\,(\ref{psiStructure}) and (\ref{chiStructure}) are evaluated, Eq.\,(\ref{uLandauLifshitz}) forms an analytic solution of the LL Eq.\,(\ref{LLEq}) for the transverse wave potential Eq.\,(\ref{WavePotential}) in a closed form.  
The equivalence of the solution presented here with~\cite{DiPiazza} is proven analytically in Appendix~\ref{app:solncomp}.  A direct comparison of numerical results for the physical situation studied in~\cite{DiPiazza2} is also presented there.

Since the LF Eq.\,(\ref{LorentzEq}) is obtained from the LL Eq.\,(\ref{LLEq}) in the limit $\tau_0\rightarrow 0$ and the field is continuous, the solution of the LL equation contains the solution of the LF Eq.\,(\ref{LorentzEq}) as a special case. In the limit of $\tau_0\rightarrow 0$, we see that the solution Eq.\,(\ref{uLandauLifshitz}) reduces to
\begin{eqnarray} \label{uLorentz}
u^\alpha&=&u^\alpha_0+a_0 (\hat{A}^\alpha-\hat{A}^\alpha_0) \\ \notag
&&-\frac{k^\alpha}{k \cdot u_0}\Bigg[ a_0 (\hat{A}-\hat{A}_0)\cdot u_0+a_0 ^2 \frac{(\hat{A}-\hat{A}_0)^2}{2}\Bigg] ,
\end{eqnarray}
which is the known solution~\cite{Sarachik:1970ap} to the LF Eq.\,(\ref{LorentzEq}). In this case $\xi$ and $\tau$ have linear dependence, and  Eq.\,(\ref{XiTauLL}) reduces to
\beqn \label{xiTauLorentz}
\xi(\tau)=\left(k \cdot u_0\right) \tau.
\eeqn

Comparing the solution Eq.\,(\ref{uLandauLifshitz}) to the LL equation with the solution Eq.\,(\ref{uLorentz}) to the LF equation we see that $\tau_0$ and $\tau_0^2$ appear in the former.  Terms that are quadratic in $\tau_0$ are in the direction of the wave vector $k^\alpha$. The linear terms in $\tau_0$ were separated intentionally in Eq.\,(\ref{uLandauLifshitz}) to terms in the direction of $\varepsilon^\alpha$ (last term in the first line) and terms in the direction of $k^\alpha$ (whole second line).  Most of the new terms that produce the deviation from the LF prediction are in the direction of propagation of the wave, as one would expect. 

Despite the appearance of (at most) quadratic terms in the solution Eq.\,(\ref{uLandauLifshitz}), it is important to realize that the dependence of $u^\alpha$ on $\tau_0$ is more complicated than what meets the eye. The projection $k \cdot u$ can be expanded in an (infinite) Taylor series in $\tau_0$ using Eq.\,(\ref{ku}), from which it is seen that $u^\alpha$ contains an infinite number of powers of $\tau_0$.  This all-orders expansion does not imply that the LL equation gives an exact description of the radiating charged particle. The LL Eq.\,(\ref{uLandauLifshitz}) is an approximation taking into account the radiative energy loss of the particle only from the zeroth order (Lorentz) motion given in Eq.\,(\ref{LorentzEq}). For a more complete discussion of the physics and assumptions involved in this expansion, see~\cite{Gralla:2009md}.

The solution Eq.\,(\ref{uLandauLifshitz}) to the LL equation contains powers in the intensity up to $a_0^6$,  four powers higher than $a_0^2$, the highest power appearing in the solution Eq.\,(\ref{uLorentz}) to the LF equation. Consequently, increasing the intensity of the electromagnetic wave brings the dynamics closer to the RR dominated regime, in which the predictions of the LL model differ significantly from those of the LF equation as we now discuss.

\section{Radiation Emission} \label{RadiationSection}
\subsection{Radiation reaction}
The rate of energy-momentum loss due to radiation is normally computed 
by the Abraham-Heaviside formula
\beqn \label{AbrahamHeaviside}
\frac{dP^{\alpha}}{d\tau}=-\frac{2}{3} e^{2}\left(\dot{u}^2\right)u^{\alpha}.
\eeqn
However, a careful examination of its derivation~\cite{Jackson} 
shows that Eq.\,(\ref{AbrahamHeaviside}) depends on the equation of motion, 
which is assumed to be the LF Eq.\,(\ref{LorentzEq}). 
This renders a serious difficulty in the computation of radiation 
emission for the LL equation, as it currently does not have a corresponding 
expression for the rate of energy lost in radiation.  The difficulty stems from the implicit description of radiation reaction in terms of the Maxwell equations and thus absence of a complete theory of charged-particle dynamics, as would be encoded in a single unified action principle.
We are therefore compelled to proceed by retaining the usual 
expressions for radiation emission obtained from the Maxwell-Lorentz theory. 
The present work should be considered as a case study for the radiation emission 
where RR effects are present, computed within the 
limitations of the current theoretical framework, as no other approach yet exists.

The rate of energy-momentum loss Eq.\,(\ref{AbrahamHeaviside}) is a four-vector that requires the knowledge of the four-velocity and four-acceleration of the particle. The four-velocity was already given in Eq.\,(\ref{uLandauLifshitz}), and the four-acceleration is supplied in Appendix~\ref{4acc-formulae}. Since the four-velocity and four-acceleration are different in the Lorentz and in the LL dynamics, the radiation emission produces different results as well. The rate (with respect to proper time) at which energy is emitted from the particle is just the $0$-component of the four-vector in Eq.\,(\ref{AbrahamHeaviside}). In order to obtain $\mathcal{R}$, the rate at which energy is emitted with respect to {\it laboratory time}, we notice that
\beqn \label{Radiation2}
\mathcal{R} \equiv \frac{dP^0}{dt}= \frac{d\tau}{dt} \frac{dP^0}{d\tau}= -\frac{2}{3} e^2 \dot{u}^2,
\eeqn
where we used Eq.\,(\ref{AbrahamHeaviside}) in the last equality. This is nothing other than Eq.\,(\ref{Radiation}) which was stated earlier in the introduction. Evaluating $\mathcal{R}$ using the four-acceleration Eq.\,\eqref{aLandauLifshitz} yields the rate at which energy is radiated from the charge with respect to the laboratory time according to the LL equation
\begin{eqnarray} \label{RadiationLandauLifshitz} \notag
\mathcal{R}&=&-\frac{2}{3} e^2 \frac{(k \cdot u)^4}{(k \cdot u_0)^2} \left\{a_0^2 \hat{A}'^2 \right. \\ \notag
&& \left. + \tau_0 (k \cdot u_0) \left[2a_0^2 \hat{A}''\cdot\hat{A}'-2a_0^4\psi \hat{A}'^2\right] \right. \\ \notag
&& \left. + \tau_0^2 (k\cdot u_0)^2 \bigg[ a_0^2 \hat{A}''^2-2a_0^4 \psi \hat{A}'\cdot\hat{A}''  \right. \\ \notag
&& \hspace*{0.8 cm} \left. \left. +a_0^4 (\hat{A}'^2)^2 \frac{k \cdot u}{k\cdot u_0}\left(\frac{k \cdot u}{k \cdot u_0}-2\right) + a_0^6 \psi^2 \hat{A}'^2 \right]  \right. \\ \notag
&& \left. + \tau_0^3(k\cdot u) (k \cdot u_0)^2 (\hat{A}'^2)^2 a_0^6 \psi \left(2-2\frac{k\cdot u}{k\cdot u_0}\right) \right. \\ 
&& \left. + \tau_0^4 (k\cdot u)^2 (k \cdot u_0)^2 (\hat{A}'^2)^2 a_0^8 \psi^2 \right\}.
\end{eqnarray}
With the aid of Eq.\,(\ref{ku}), it can be written as
\begin{eqnarray} \label{RadiationLandauLifshitz2}
\mathcal{R} \!&=&\! -\frac{2}{3} e^2 \frac{(k \cdot u)^4}{(k\cdot u_0)^2} \\
&&\times \bigg\{ \left[a_0 \hat{A}'+ (k\cdot u_0) \tau_0 a_0 \hat{A}'' - (k \cdot u_0) \tau_0 a_0 ^3 \psi \hat{A}' \right]^2 \notag \\
&&\hspace*{0.8cm}  -(k \cdot u_0)^2 \tau_0 ^2 a_0 ^4 (\hat{A}'^2)^2 \bigg\}.  \notag 
\end{eqnarray}
Since the four-vector in the brackets is a spacelike four-vector, this is a manifestly
positive-definite expression [this is also evident from its definition in Eq.\,(\ref{Radiation2})]. By taking the limit $\tau_0 \rightarrow 0$ one
obtains the radiation rate in the absence of RR effects
\beqn \label{RadiationLorentz}
\mathcal{R} = -\frac{2}{3} e^2 (k \cdot u_0)^2 a_0^2 \hat{A}'^2.
\eeqn

This is the rate at which energy is radiated from the particle in the case of 
motion according to the LF Eq.\,(\ref{LorentzEq}). Comparing Eq.\,(\ref{RadiationLandauLifshitz}) with Eq.\,(\ref{RadiationLorentz}) allows 
us to determine the condition in which the predictions of the LL model depart from the predictions of the LF equation. The particle has an initial velocity of $u^\alpha_0=\gamma_0 (1,\vec{v_0})$, and 
\beqn
k\cdot u_0 = \gamma_0 (\omega - \vec{k} \cdot \vec{v_0}).
\eeqn

If the wave vector $\vec{k}$ and the initial 3-velocity $v_0$ point in the same direction, $k\cdot u_0\leq \omega$. However, when the wave vector $\vec{k}$ and the initial 3-velocity $v_0$ are in opposite directions, $k\cdot u_0$ has order of magnitude $\omega \gamma_0$. This corresponds to a head-on collision between the electron and the laser beam, which thanks to the factor of $\gamma_0$ enhances the radiation-reaction effects greatly as we will see next. When $a_0 \gg 1$, the leading order correction to the LF equation will be the $\tau_0$ term, in which the dominant term is that proportional to $a_0^4$. We therefore enter the RR dominated regime when the corrections supplied by the LL equation become of the same order of magnitude as the ones in Eq.\,(\ref{RadiationLorentz}), namely as
\beqn\label{RRDRcond0}
a_0^2 \sim (\omega \tau_0) \gamma_0 a_0 ^4.
\eeqn
For an electron traveling into a laser beam with wavelength $\lambda \sim 942 \, {\rm nm}$ (or equivalently frequency $\omega \sim 2 \,{\rm fs}^{-1}$) we have $\omega \tau_0 \sim 10^{-8}$, and we enter the RR dominated regime as
\beqn \label{RRDRCondition}
\gamma_0 a_0 ^2 \sim 10^8.
\eeqn 
Obtained on the basis of an analytical solution, these conditions provide nontrivial validation of the conditions found by~\cite{LandauLifshitz} and~\cite{Koga2}.  Furthermore, with an explicit solution in hand we can go further and verify the condition \eqref{RRDRcond0} by directly comparing experimental observables predicted by the LF and LL equations.

We see that there is an inverse squared relation between the initial energy of the particle and the intensity of the laser. Notice, however, that once we are deep inside the radiation dominated regime, i.e. when
\beqn
a_0^2 \ll (\omega \tau_0) \gamma_0 a_0 ^4
\eeqn
higher order terms in $\tau_0$ begin to dominate the emission from the particle and not the term linear in $\tau_0$. This still results in a different radiation emission than the one predicted by the LF equation.

\begin{figure}[t]
\begin{center}
\includegraphics[width=3.4in,height=3.4in]{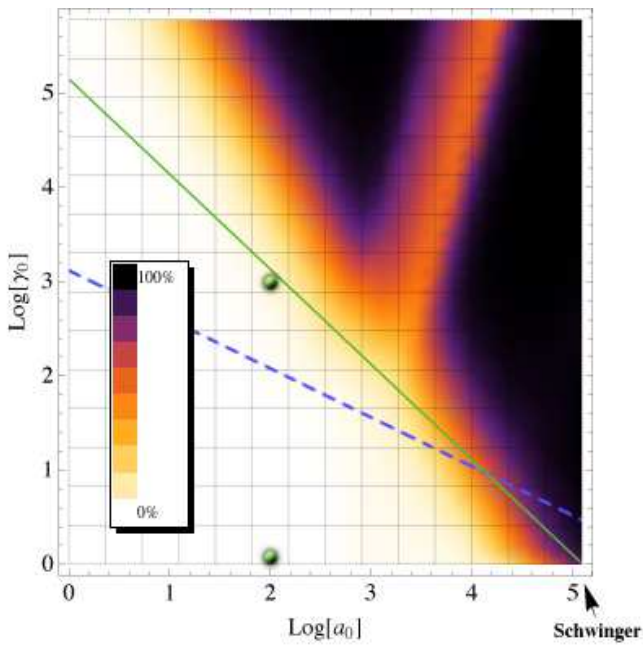}
\caption{\label{RRDRGraph} A density plot of Eq.\,(\ref{RRDRCP}) for a LP wave with wavelength $\lambda=942 \, {\rm nm}$. Above the dashed (blue) line given by Eq.\,(\ref{LLValid}), the predictions of the LL equation differ from those of the Lorentz-Abraham-Dirac equation. Along the solid green line the electron experiences a force   of order unity [$m^2$] [see Eq.\,(\ref{AccelerationUnity})].  The Schwinger critical field $E_c = m^2c^3/e\hbar$ is indicated in the bottom right corner.  The solid circles are the values that are studied in Sec.~\ref{ExamplesSection}.}
\end{center}
\end{figure}

Figure \ref{RRDRGraph} demonstrates the validity of Eq.\,(\ref{RRDRCondition}) in the case of a head-on collision between an electron and a linearly polarized plane wave. The density at each value of the intensity $a_0$ and $\gamma_0$ was computed numerically over one period of the wave (in $\xi$) by considering the statistical deviation in the energy of the particle predicted by the LL equation and the LF equation. It is a value between $0$ and $1$ given by
\beqn \label{RRDRCP}
\Delta=\frac{1}{2 \pi} \int_0 ^{2\pi} \frac{| E_{LL}(\xi)-E_{\rm Lorentz}(\xi) |}{|E_{LL}(\xi)+E_{\rm Lorentz}(\xi)|} d\xi.
\eeqn
 The result is striking, as Fig.~\ref{RRDRGraph} shows that already {\it with current laser systems RR effects can be tested}. In fact, the figure shows that RR effects can already be tested for high-intensity lasers with $a_0=100$ and $\gamma_0=1000$ (a ``modest'' initial energy of $E_0 = 0.511\,\rm{GeV}$), as is demonstrated in the next section. The boundary of the shaded domain 
is in agreement with condition Eq.\,(\ref{RRDRCondition}).

Recall that the LL Eq.\,(\ref{LLEq}) was derived from the LAD Eq. (\ref{LADEq}) using perturbation theory \cite{LandauLifshitz}.  The earlier edition \cite{LandauLifshitz2ed} makes it clear that the radiation-reaction forces of the LAD and LL equations agree (perturbatively) when the Lorentz force dominates.  
However, solutions to the LAD or LL equation need not be equivalent in general, though a relationship can be established in the case that the radiation-reaction force remains subdominant~\cite{Spohn:1999uf}.

Specifically, LAD and LL predictions can disagree when radiation-reaction forces dominate the Lorentz force.  In fact, a difference is possible in a region of parameter space where classical dynamics are valid, as is seen by comparing the dominant term in the LL force~Eq.\,(\ref{LLEq}) 
\beqn
-\frac{e^2 \tau_0}{m} F^{\beta\gamma} F_{\gamma\delta} u^\delta u_\beta u^\alpha \sim - \tau_0 m \omega ^2 a_0 ^2 \gamma_0 ^3,
\eeqn
to the LF, 
\beqn
-e F^{\alpha\beta} u_\beta \sim -m \omega a_0 \gamma_0
\eeqn
which gives 
\beqn \label{LLValid0}
\tau_0 m \omega ^2 a_0 ^2 \gamma_0 ^3 \ll m \omega a_0 \gamma_0.
\eeqn
For wavelengths $\lambda \sim 942 \,{\rm nm}$ we have $\omega \tau_0 \sim 10^{-8}$, and the LL equation is compatible with the LAD equation when
\beqn \label{LLValid}
a_0 \gamma_0 ^2 \ll 10^8.
\eeqn 
As noted also in Sec.~76 of~\cite{LandauLifshitz}, this condition differs from Eq.\,\eqref{RRDRCondition}:
In the former, $a_0$ is proportional to the square of the inverse of $\gamma_0$ and in the latter $\gamma_0$ is proportional to the square of the inverse of $a_0$.  Consequently, there are values of $a_0$ and $\gamma_0$ for which radiation reaction dominates the dynamics.

We can therefore arrange experiments for which the LF Eq.\,(\ref{LorentzEq}), the LAD Eq.\,(\ref{LADEq}) and the LL Eq.\,(\ref{LLEq}) may {\it all predict different particle dynamics}.  For these values we cannot only measure RR effects, but can also distinguish the LL equation from the LAD equation.  The dashed (blue) line in Fig.~\ref{RRDRGraph}  represents the quantitative condition in Eq.\,(\ref{LLValid}), for values at which the RR force is $1\%$ the magnitude of the LF, and above it, LL and LAD dynamics will be distinguishable.  For example, the point marked at intensity $a_0=100$, initial $\gamma_0\sim 10^3$ suffices to probe the RR dominated regime in which the LL Eq.\,(\ref{LLEq}) is incompatible with the LAD Eq.\:(\ref{LADEq}).

Since the RR force stems from radiation emission, and radiation is emitted
only if the particle is being accelerated [see Eq.\,(\ref{Radiation})], the real 
key physical property that is responsible for the RR is the acceleration. 
In natural units, the magnitude of acceleration $[m]$ and force $[m^2]$
obtained by hitting a resting electron with a laser beam
is usually not approaching unity (measured in units of the electron's mass).
For example, for $a_0=100$ the maximal Lorentz invariant force 
is only $\sqrt{-\dot{u}^\alpha \dot{u}_\alpha} = 8\times 10^{-4} \, {\rm m_e}$.

\begin{figure}[t]
\begin{center}
\includegraphics[scale=0.93]{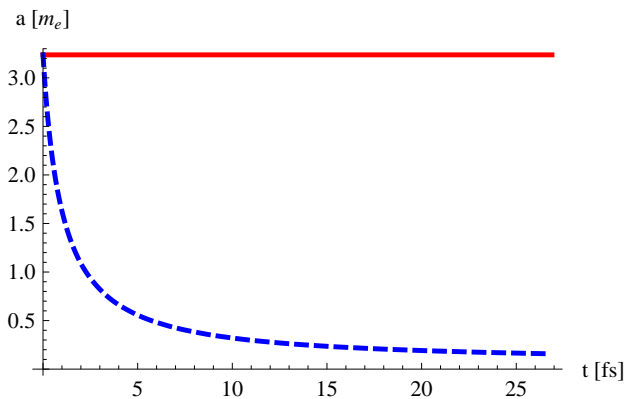}
\caption{\label{CPAcceleration} The Lorentz invariant acceleration $\sqrt{-\dot{u}^\alpha \dot{u}_\alpha}$ in natural units for a circularly polarized  laser wave with $a_0=100$ and initial $\gamma_0=1000$.  The solid red line is the acceleration in the LF case, while the dashed blue line gives the acceleration according to the LL.}
\end{center}
\end{figure}

However, in a head-on collision between the electron and the laser pulse,
the ``critical'' force (i.e. acceleration of the order of the electron mass)
can be achieved easily, as is presented in Fig.~\ref{CPAcceleration}. 
This figure shows an acceleration of order unity, achieved by colliding 
a relativistic electron ($\gamma_0=1000$) with a laser with intensity 
$a_0=100$. In fact, by computing the LF we see that ``unity'' acceleration is achieved
when
\beqn \label{AccelerationUnity}
\omega a_0 \gamma_0 \sim m_e,
\eeqn
corresponding to the appearance of the ``critical'' field strength $E_c=m^2c^3/e\hbar$ in the electron rest frame.  This defines the solid (green) line in Fig.~\ref{RRDRGraph}. Quantum effects may become relevant beyond this boundary~\cite{Ritus1985,Baier:1998vh}, and we have chosen examples within the domain where classical dynamics certainly dominate.  It is noteworthy that Eq.\,\eqref{AccelerationUnity} also loosely demarcates the domain of validity of the LL equation itself, i.e., the lighter areas of Fig.~\ref{RRDRGraph} where the predicted radiation remains ``small.''
For particles other than the electron, condition Eq.\,(\ref{AccelerationUnity}) differs, as the constant $a_0$ defined in Eq.\,(\ref{a_0}) depends on the mass of the particle.  

Figure~\ref{RRDRGraph} exhibits the existence of an area of $(a_0,\gamma_0)$ parameter space especially sensitive for experiments probing radiation reaction.  Between the solid (green) and dashed (blue) lines, the classical LAD and LL equations are both valid, but their dynamical predictions in general can differ observably.  The importance of this recognition is the reason we have chosen to provide in Sec.~\ref{ExamplesSection} numerical solutions for the point marked at $(a_0=100,\gamma_0=1000)$.

\subsection{Angular Distribution of Radiation}
The energy flux measured in the laboratory frame is given by the Poynting vector ~\cite{Jackson}
\beqn \label{PoyntingVector}
\vec{S} = \frac{1}{4\pi} \vec{E}\times\vec{B},
\eeqn
where $\vec{E},\vec{B}$ are the electric and magnetic field respectively.
Let $y^\alpha=\left(t,\vec{y}\right)$ be the spacetime point at which we evaluate
the electromagnetic field, and denote by $\tau_0$ the proper time at which the 
worldline of the charge $x^\alpha$ intersects the past light cone emanating
from $y^\alpha$. The retarded time $\tau_0$ is defined by the light-cone constraint
\beqn \label{LightConeConstraint}
t - x_0(\tau_0) = R,
\eeqn
where we denote by $R$ the spatial separation between the two events, namely
\beqn
R=\left| \vec{y}-\vec{x}(\tau_0) \right|.
\eeqn

Since the Poynting vector in Eq.\,(\ref{PoyntingVector}) is quadratic in the fields,
and radiation is defined as the energy transported to infinity, only fields that are
falling off as $R^{-1}$ or slower radiate. Consequently, it can be 
shown~\cite{Jackson} that the energy flux radiated to infinity by a single charge is
\beqn \label{RadialPoynting}
\left[\vec{S}\cdot \vec n\right]_\text{ret} = \frac{e^2}{4\pi} \Bigg\{\frac{1}{R^2} \left| \frac{\vec{n}\times\left[\left(\vec{n}-\vec{v}\right)\times\vec{a}\right]}{\left(1-\vec{v}\cdot\vec{n}\right)^3} \right| ^2 \Bigg\}_\text{ret} ,
\eeqn
where $\vec{n}$ is a unit vector in the direction of $\vec{y}-\vec{x(\tau)}$,
$\vec{v},\vec{a}$ are the three-dimensional velocity and acceleration of the
charge respectively, and subscript $\text{ret}$ means that the quantity in
the brackets is to be evaluated at the retarded time $\tau_0$.

The total amount of energy per unit solid angle emitted during a finite acceleration period $T$ is obtained by integrating the Poynting vector from the retarded lab time $t=0$ to $t=T$ and is given by ~\cite{Jackson}
\beqn \label{RadiationDistributionGeneral}
\frac{dE}{d\Omega} = \int_0 ^T R^2 (\vec{S}\cdot\vec{n}) (1 - \vec{v}\cdot\vec{n}) dt.
\eeqn
Since the solution in Eq.\,(\ref{uLandauLifshitz}) is given in terms of $\xi$, it is useful to express the integrand using the variable $\xi$ as well, as
\begin{eqnarray} \label{RadiationDistribution}
\frac{dE}{d\Omega} &=& \int_0 ^{\xi(t=T)} R^2 (\vec{S}\cdot\vec{n}) (1 - \vec{v}\cdot\vec{n}) \frac{dt}{d\tau} \frac{d\tau}{d \xi} d\xi \\ \notag
&=& \frac{e^2}{4\pi} \int_0 ^{\xi(t=T)} \frac{\left|\vec{n}\times\left[\left(\vec{n}-\vec{v}\right)\times\vec{a}\right]^2\right|}{\left(1-\vec{v}\cdot\vec{n}\right)^5} \frac{\gamma}{k \cdot u} d\xi.
\end{eqnarray}

In order to compute Eq.\,(\ref{RadiationDistribution}) we choose the coordinate system such that the electromagnetic wave propagates in the positive $z$ direction while the particle initially moves in the negative $z$ direction. Equation (\ref{RadiationDistribution}) results in a cumbersome expression in the general case, which does not add much to the understanding of the physics of the problem. Instead, in the next section we will consider the angular distribution of the energy emitted in the specific cases presented earlier, of linearly and circularly polarized plane waves.

\begin{figure}
\begin{center}
\includegraphics[scale=0.95]{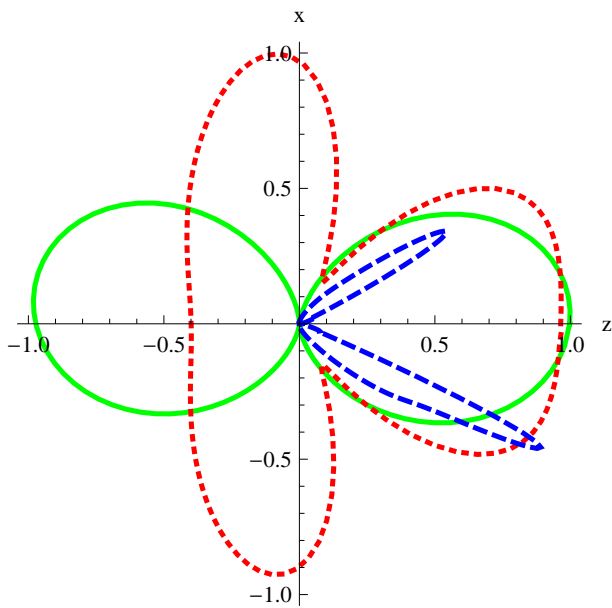}
\caption{\label{RadiationGraph} The angular distribution of radiation for a linearly polarized wave. This is the normalized radiation distribution for an electron initially at rest, after interacting with a laser with $a_0=0.1$, $a_0=1$ and $a_0=10$ plotted in solid green, dotted red and dashed blue lines respectively. This plot is identical for the LF and for the LL equation.}
\end{center}
\end{figure}

Figure \ref{RadiationGraph} shows the normalized angular distribution of radiation in the $x-z$ plane in the laboratory frame for a LP plane wave hitting an electron initially at rest. The distribution was normalized to unity by dividing it by its maximal value in order to demonstrate the different shapes of radiation distribution for a laser beam with wavelength $\lambda=942 \, {\rm nm}$ and intensity $a_0=0.1,1,10$. For these intensities (and an electron at rest) the LF equation and the LL equation produce indistinguishable radiation distribution. The asymmetry in the dashed blue curve for $a_0=10$ is because the plot was generated for a laser wave duration of $26.8\,{\rm fs}$. For these values of the wave duration and laser intensity, the velocity of the electron in the $x$ direction does not finish an integer number of periods and results in an asymmetric motion in the $x$ direction and consequently in the angular distribution of radiation.

\section{Examples} \label{ExamplesSection}
We now discuss our main examples of an electron traveling in linearly and circularly polarized plane waves. Assume that in the lab frame, the electron is initially traveling towards the electromagnetic wave to a head-on photon-electron collision at time $t=0$. For the figures and the numerical values, we take an electromagnetic  wave with intensity $a_0=100$, wavelength $\lambda=942 \, {\rm nm}$, frequency $\omega=2 \,{\rm fs}^{-1}$ and a pulse that runs for a duration of $26.8 \,{\rm fs}$. The RR effects are demonstrated graphically for an electron initially at rest $\gamma_0=1$ and a highly relativistic electron with initial $\gamma_0=1000$.

\subsection{Linearly polarized plane wave}
The definitions given in Eq.\,(\ref{LPPlaneWave}) give the structure integrals Eqs.\,(\ref{psiStructure}) and (\ref{chiStructure})
\begin{eqnarray}
\psi &=& -\frac{1}{2}(\xi+\sin\xi \cos\xi) \\ \notag
\chi^\alpha &=& \left(0,\frac{1}{6}(2-3\cos\xi-3\xi\sin\xi+\cos^3 \xi),0,0\right).
\end{eqnarray}
The projection of the four-velocity onto the wave vector is given by Eq.\,(\ref{ku}) and yields
\beqn
k \cdot u = \frac{k \cdot u_0}{1+\frac{1}{4} \tau_0 a_0^2 (k \cdot u_0) (2\xi+\sin2\xi)}.
\eeqn
In Appendix~\ref{4vel-formulae} we provide the explicit expression for the four-velocity of the particle in this case, as is given by Eq.\,(\ref{uLandauLifshitz}).

Equation\:(\ref{XiTauLL}) gives the relationship between $\xi$ and $\tau$, which is in this case
\beqn \label{XiTauLP}
\tau(\xi)=\frac{\xi}{k\cdot u_0} + \frac{1}{8}a_0^2 \tau_0 \left[1+2\xi^2-\cos2\xi\right].
\eeqn
Notice that the limit $\tau_0\rightarrow 0$ indeed exists, where we obtain the linear relation of Eq.\,(\ref{xiTauLorentz}). In fact, in this particular case of the LF equation with a linearly polarized plane wave, Eq.\,(\ref{xiTauLorentz}) is a known result~\cite{Sarachik:1970ap}, which means that the four-acceleration of the particle is always orthogonal to the wave vector $k^\alpha$ in the absence of RR effects. 

\begin{figure}
\begin{center}
\includegraphics[scale=0.95]{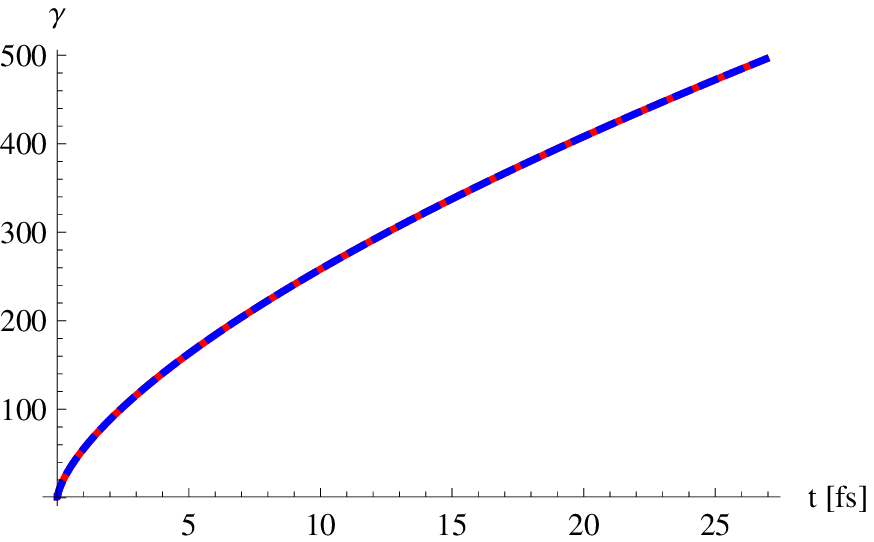}
\includegraphics[scale=0.95]{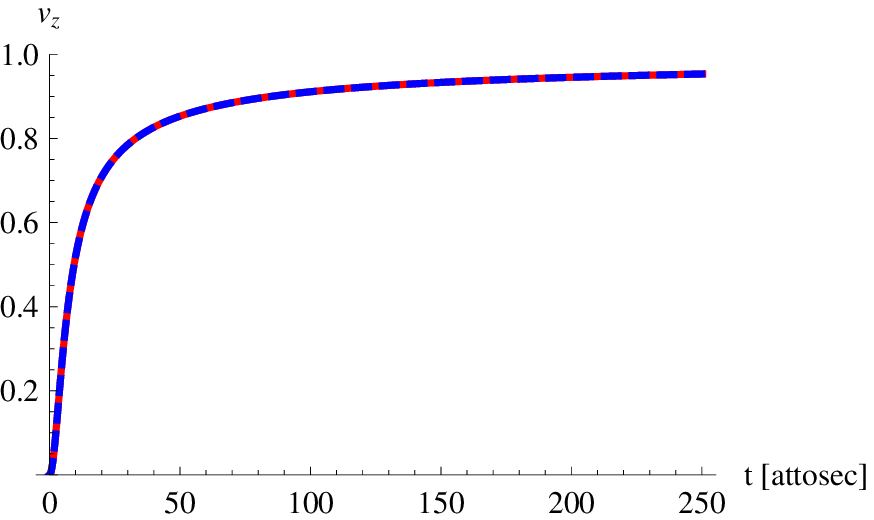}
\caption{\label{GraphsTGamma1a100Graphs} The electron's Lorentz factor $\gamma$ and the longitudinal velocity $v_z$ for a linearly polarized wave with $a_0=100$ and wavelength $\lambda=942\,{\rm nm}$ hitting an electron initially at rest. The solid red line represents the solution of the LF equation, while the dashed blue line represents the LL equation which is indistinguishable in this case.}
\end{center}
\end{figure}

Figure \ref{GraphsTGamma1a100Graphs} shows $\gamma$ and the longitudinal velocity for an electron that is initially at rest. All figures are presented in the laboratory frame with respect to the laboratory time $t$. As Fig.~\ref{RRDRGraph} shows, this is still far from the RR dominated regime and the predictions of the LL equation are practically indistinguishable from the ones given by the LF equation. The particle is boosted rapidly in the direction of the laser wave $\hat{z}$ and is relativistic after approximately $100 \, {\rm as}$.

\begin{figure}[t]
\begin{center}
\includegraphics[scale=0.95]{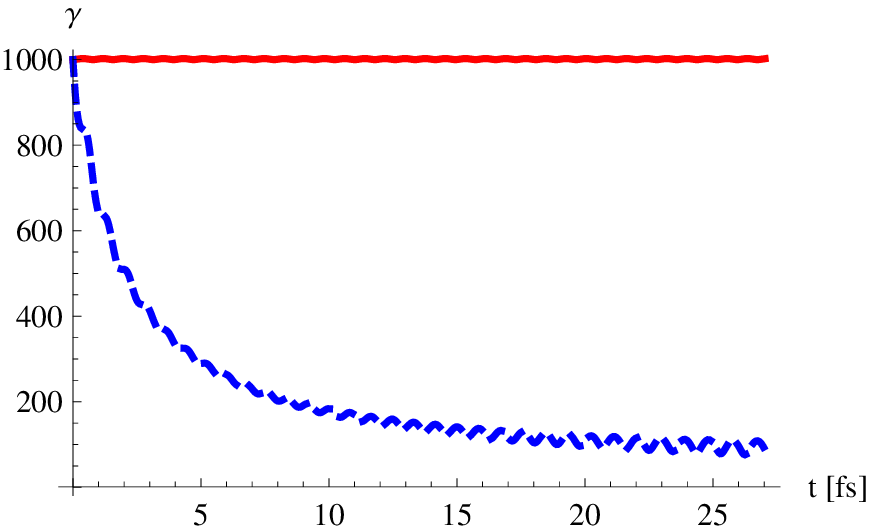}
\includegraphics[scale=0.95]{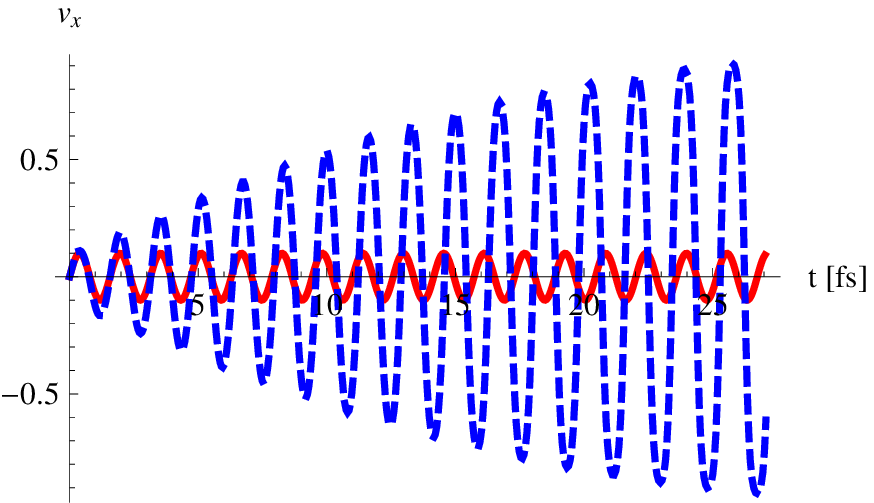}
\includegraphics[scale=0.95]{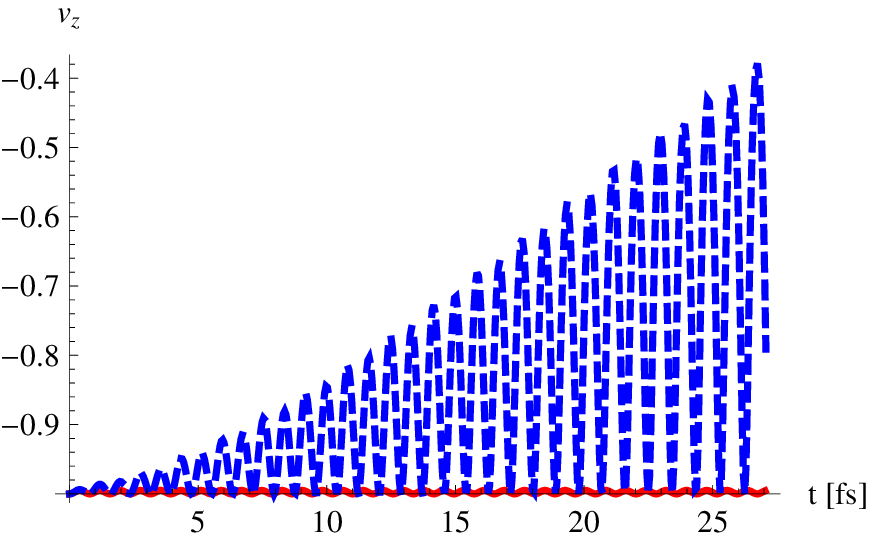}
\caption{\label{LPGraphsTGamma1000a100Graphs} The electron's Lorentz factor $\gamma$, the transverse velocity $v_x$ and the longitudinal velocity $v_z$ in the laboratory frame as a function of lab time for a linearly polarized  wave with $a_0=100$ and an electron with initial $\gamma_0=1000$. The solid red line represents the solution of the LF equation, while the blue (dashed) line represents the LL equation.}
\end{center}
\end{figure}

If the particle is initially moving in the negative $\hat{z}$ direction, there is a head-on collision between the wave and the particle. An electron with initial $\gamma_0=1000$ ($E_0=0.511 \,{\rm GeV}$) is on the critical line of the RR dominated regime. It is presented in Fig.~\ref{LPGraphsTGamma1000a100Graphs}. The pulse duration is not long enough to flip the direction of the highly relativistic electron (a $180\, {\rm fs}$ pulse duration is needed). The radiation emitted by the particle slows the electron down rapidly in the direction of the wave, while it gains momentum in the direction of polarization. This is only true in the case of the LL model (the dashed blue line) due to the RR effects. In the absence of RR (the solid red line) the pulse is incapable of slowing the particle down.

\begin{figure}[t]
\begin{center}
\includegraphics[scale=0.95]{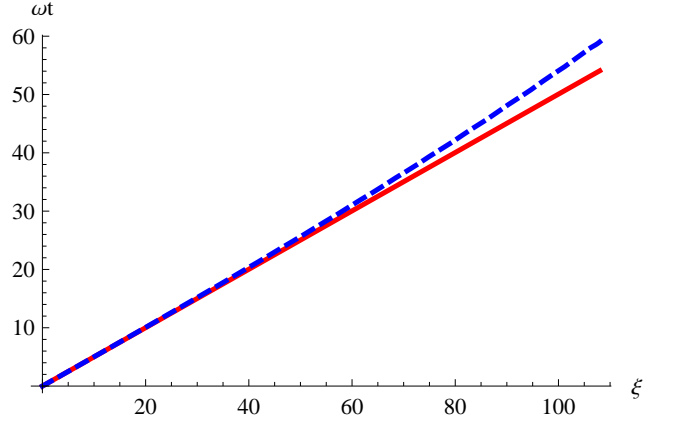}
\caption{\label{LPGraphsTXiGamma1000a100} The dependence between the laboratory time $t$ and the variable $\xi$ in a LP plane wave for LF (solid red line) and LL (dashed blue line) dynamics. The duration of $\xi$ is doubled from the original duration of the wave in $\omega t$. }
\end{center}
\end{figure}

Although the wave is active for about $8$ periods, we see from Fig.~\ref{LPGraphsTGamma1000a100Graphs} that the transverse velocity has $16$ periods, while the longitudinal velocity has $32$ periods. The reason for this effect is twofold. First, recall that the solution Eq.\,(\ref{uLandauLifshitz}) was given in terms of the variable $\xi$, which has a nontrivial relation to the laboratory time in which we measure the velocities of the particle. This relation is demonstrated in Fig.~\ref{LPGraphsTXiGamma1000a100} in which we see a doubling of the duration for $\xi$. Second, a careful examination of the solution for a linearly polarized plane wave (in Appendix~\ref{4vel-formulae}) shows that the period is doubled for the longitudinal velocity once more, resulting in $32$ periods. The second effect is a feature of the linear polarization and is absent in the case of a circularly polarized wave.

\begin{figure}
\begin{center}
\includegraphics[scale=0.95]{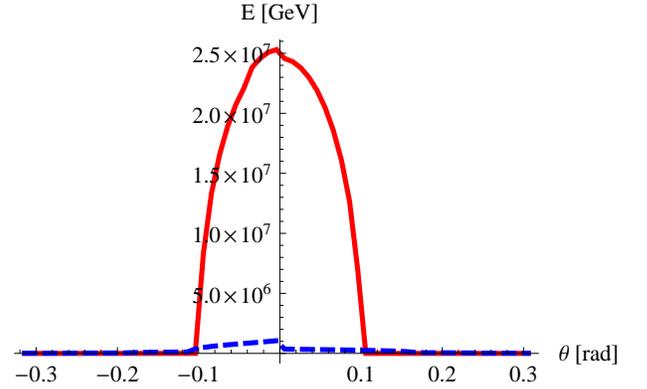}
\caption{\label{LPRadiationGraph} Radiation emission for a linearly polarized wave with $a_0=100$ colliding head-on with
an electron with initial $\gamma_0=1000$. The angle $\theta$ is measured on the $x-z$ plane, starting from the negative $z$ axis. The radiation for the LF equation (solid red line) is 1 order of magnitude
greater than the radiation in the case of the LL equation (the dashed blue line).}
\end{center}
\end{figure}

Since the electron is highly relativistic, the radiation is emitted almost purely in the direction of its initial motion,
namely. in the negative $z$ direction. The different radiation emission from the LF equation and the LL equation is shown
in Fig.~\ref{LPRadiationGraph} where we see that the LF dynamics produce different radiation emission
than the LL dynamics.

\subsection{Circularly polarized plane wave}

We use Eq.\,(\ref{CPPlaneWave}) to evaluate the structure integrals Eqs.\,(\ref{psiStructure}) and (\ref{chiStructure})
\begin{eqnarray}
\psi &=& -\xi \\ \notag
\chi^\alpha &=& (0,\sin(\xi)-\xi\cos(\xi),1-\cos(\xi)-\xi\sin(\xi),0).
\end{eqnarray}
The projection of the four-velocity onto the wave vector is given by Eq.\,(\ref{ku}) and yields
\beqn
k \cdot u = \frac{k \cdot u_0}{1+\tau_0 a_0^2 (k \cdot u_0) \xi}.
\eeqn
The four-velocity of the particle is given by Eq.\,(\ref{uLandauLifshitz}) and the analytic expression for it in this case is given in Appendix~\ref{4vel-formulae}. In order to have the four-velocity in terms of the proper time $\tau$, we need to invert Eq.\,(\ref{XiTauLL}). In the case of a circularly polarized plane wave, we can find the inverse explicitly and obtain
\beqn \label{TauXiCP}
\xi(\tau)=\frac{\sqrt{1+2a_0^2 (k \cdot u_0)^2 \tau_0 \tau}-1}{a_0 ^2 (k \cdot u_0) \tau_0}.
\eeqn

Since a circularly polarized wave with intensity $a_0=100$ produces similar 
results to a linearly polarized plane wave of the same intensity
(Fig.~\ref{GraphsTGamma1a100Graphs}), the graphs are not presented here.
However, because for the circularly polarized case in Eq.\,(\ref{CPPlaneWave}) there is no choice
of phase $\xi_0$ of the wave for which the transverse motion of the resting electron vanishes,
the phase at the instant of collision has physical significance resulting
in a drift in the direction of initial polarization ($\hat{x}$ in this case).
This drift is a nonphysical consequence of the nonphysical electromagnetic field.
A different choice of initial phase for the linearly polarized wave produces a drift in 
the direction of polarization as well. This mathematical artifact is a result of 
a plane wave whose duration is infinitely long, as one can see from Eq.\,(\ref{uLandauLifshitz})
that the solution depends explicitly on the four-potential at the initial time $\hat{A}^\alpha_0$.
If the initial four-potential $\hat{A}^\alpha_0$ is nonzero, the second term 
in Eq.\,(\ref{uLandauLifshitz}) causes a drift in the opposite direction of $\hat{A}^\alpha_0$.

In practice, any experiment involving laser beams will require a certain finite duration 
on which the wave is being turned on. This can be modeled into the solution by
defining a turn-on function
\beqn \label{TurnOnFunction}
H(\xi) = \begin{cases} \sin^2(\xi/\sigma) & \xi \leq \sigma \frac{\pi}{2} \\ 1 & \xi>\sigma\frac{\pi}{2} \end{cases},
\eeqn
which vanishes for negative $\xi$, and increases continuously to $1$ over a finite period determined by the parameter $\sigma$.
The new shape function for the four-vector potential is
\beqn \label{CPPlaneWave2}
f(\xi) = \int_0 ^\xi H(y) \sqrt{2} e^{i (y)} dy
\eeqn
for the circularly polarized wave [cf. with Eq (\ref{CPPlaneWave})]. The introduction of the turn-on
function $H(\xi)$ in the shape function guarantees that the electromagnetic field will be turned on
smoothly. The integration appears since $f(\xi)$ defines the four-potential, but we would like a
smooth turn-on of the electromagnetic field which is defined in terms of the derivative of $f(\xi)$ [see Eqs.\,\eqref{WavePotential} and \eqref{PlaneWaveField}].

In the following graphs we choose the parameter $\sigma$ to be such that the electromagnetic
field is turned on on a period of about a quarter of a wavelength $\sim \lambda/4=210 \, {\rm nm}$.
Since the turn-on function makes the mathematical expressions in the solution
of Sec.~\ref{SolutionSection} cumbersome, they were omitted.

\begin{figure}[t]
\begin{center}
\includegraphics[scale=0.92]{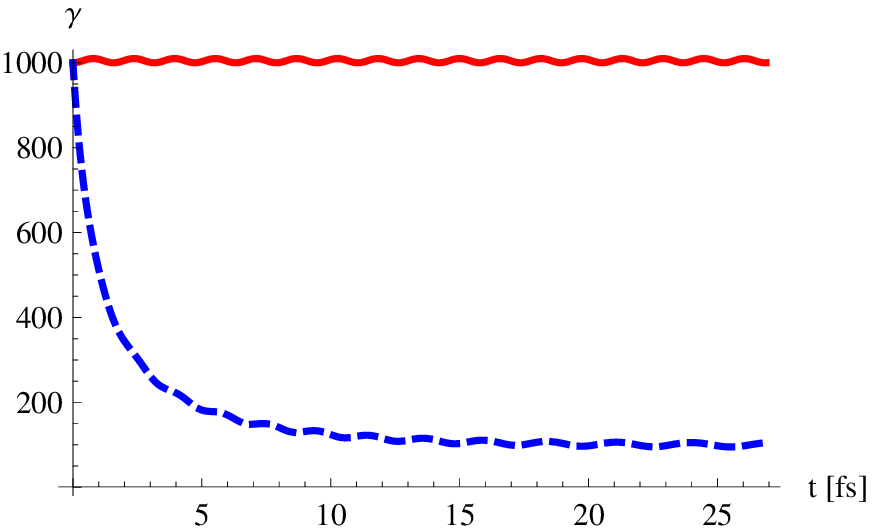}
\includegraphics[scale=0.95]{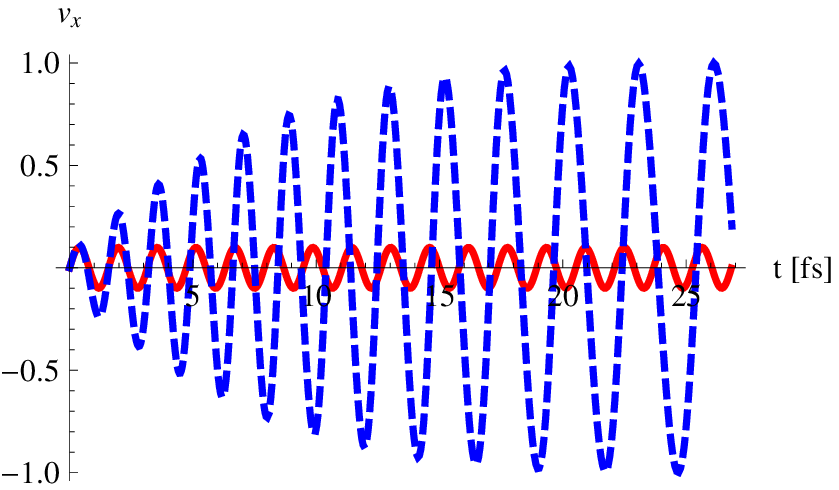}
\includegraphics[scale=0.92]{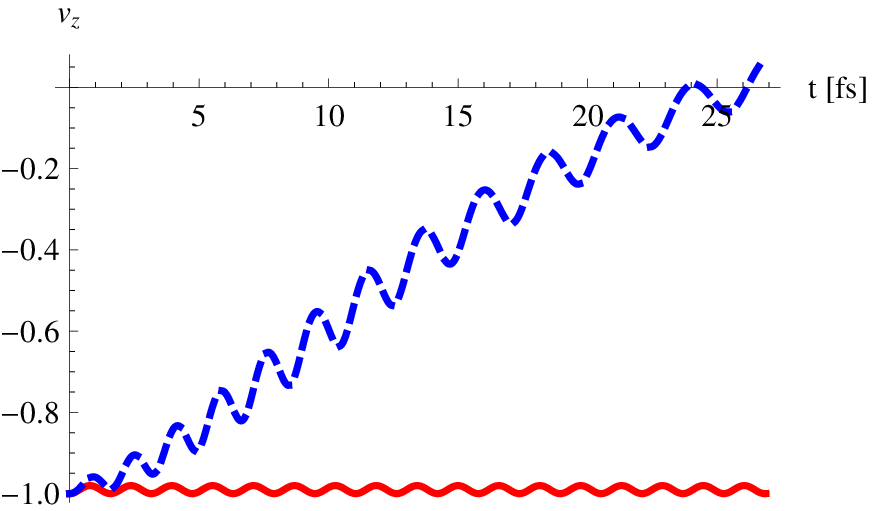}
\caption{\label{CPGraphsTGamma1000a100Graphs} The electron's Lorentz factor $\gamma$, the transverse velocity $v_x$ and the longitudinal velocity $v_z$ for a circularly polarized  wave with $a_0=100$ hitting an electron with $\gamma_0=1000$. The solid red line is the solution of the LF equation, while the dashed blue line is the LL equation.}
\end{center}
\end{figure}

\begin{figure}
\begin{center}
\includegraphics[scale=0.95]{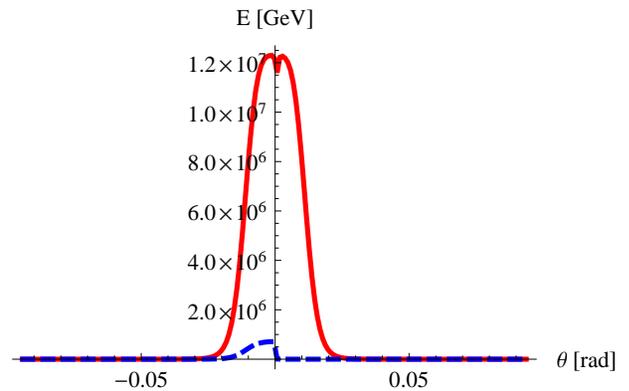}
\caption{\label{CPRadiationGraph} Radiation emission for a circularly polarized wave with $a_0=100$ colliding head-on with
an electron with initial $\gamma_0=1000$. The angle $\theta$ is measured on the $x-z$ plane, starting from the negative $z$ axis. 
The radiation for the LF equation (solid red line) is 1 order of magnitude
greater than the radiation in the case of the LL equation (the dashed blue line).}
\end{center}
\end{figure}

Figure \ref{CPGraphsTGamma1000a100Graphs} presents the head-on collision between the wave and a
relativistic particle ($\gamma_0=1000$).
The circular polarization is much more efficient in slowing the particle down, as can be seen by comparing the velocity
in the $\hat{z}$ direction in the two cases (compare Figs.~\ref{CPGraphsTGamma1000a100Graphs} and \ref{LPGraphsTGamma1000a100Graphs}).
About $26 \,{\rm fs}$ suffice for a circularly polarized wave to stop the highly relativistic electron, after which the electron
reverses its direction of motion to coincide with the direction of propagation of the wave. Figure \ref{CPRadiationGraph} presents
the radiation emission, where we see that the LF equation results in the familiar dipole radiation (the angle of separation
between the two branches of the dipole is of order $1/\gamma \sim 10^{-3}$). Similarly to the LP case, the LL equation produces radiation
which is 1 order of magnitude smaller than the LF equation.

\section{Conclusions} \label{ConclusionsSection}
The dynamics of a radiating charged particle has been the subject of 
investigation for more than a century. Lorentz himself recognized
the need to expand the Maxwell-Lorentz framework to include an additional
force due to radiation reaction. 
In the intervening years, many others sought physical consistency by
including a radiation-reaction force as a modification of the Lorentz force equation.
These efforts have produced numerous radiation-reaction models
but unlike the Maxwell-Lorentz equations none of these has been derived from an action principle.

So far, the theoretical efforts have remained experimentally untested.
Our objective here has been to connect theory with possible experiment.
We found that when colliding relativistic electrons ($\gamma\ge 1000$) with an extremely intense
laser beam ($a_0\ge 100$, power $\ge 2\times10^{22}$ W/cm$^2$), the laser 
is capable of stopping the relativistic electron once radiation-reaction force is included.

For the linearly polarized wave, the chosen pulse length was insufficient to reverse the initial momentum of the electron.  
However, a longer laser pulse could reverse the net momentum of the electron, accelerating it back in the direction from which it 
came.  Thus being a dissipative effect, radiation reaction makes the gross consequences sensitive to the 
duration of the interaction~\cite{Koga2}.  A more thorough exploration of the outcomes and observables for 
various pulse lengths using the solutions here presented would be an important preliminary to any 
experiment.

We have presented an analytic solution of the Landau-Lifshitz Eq.\,(\ref{LLEq}),
and an analysis of the radiation distribution emitted by the charged
particle. The {\it radiation-reaction-dominated 
regime criterion} given by Eq.\,(\ref{RRDRCondition}) is visualized in Fig.~\ref{RRDRGraph}, 
showing that the radiation-reaction force can be probed using current 
pulsed laser systems.  Figure \ref{RRDRGraph} exhibits the domain in which the classical dynamics here solved dominates, but the Landau-Lifshitz equation results in different radiation patterns than those obtained from the
Lorentz force Eq.\,(\ref{LorentzEq}). This suggests that a study of the radiation patterns
emitted in such experiments will provide a first experimental opportunity, well 
within the reach of current laser systems, in which radiation-reaction effects can be observed
and understood in the new physics domain.

Examining the radiation emission in both Figs.~\ref{LPRadiationGraph} and \ref{CPRadiationGraph}
one sees that the LL equation predicts less total integrated radiation emission than the 
LF equation. On first sight this appears to conflict with rapid, large amplitude oscillations of the electron.  However, we recall that the radiation rate is proportional to the acceleration, $\mathcal{R}\sim\dot u^2$, and Fig.~\ref{CPAcceleration} for $\sqrt{-\dot u^2}$ or Figs.~\ref{LPGraphsTGamma1000a100Graphs} and \ref{CPGraphsTGamma1000a100Graphs} for Lorentz factors exhibit how the LL equation permits the electron to decelerate while the LF does not.  One may thus expect the energy converted to radiation by the accelerated motion of the electron to decrease with time for the LL equation, but to remain constant for the LF equation.  As a result, the total integrated emitted power predicted by the Lorentz force can indeed be larger.  For this reason, one cannot evaluate the radiation produced in high-intensity laser-matter interactions according to the Lorentz force dynamics alone.

The examples here thus exhibit the more general point that accurate predictions of radiation emission require incorporating radiation-reaction effects whenever high accelerations are expected.  It should be emphasized that our radiation results are computed according to Dirac's evaluation of the radiated momentum, which leads in turn to the LAD dynamics.  On this point, an improvement on the present work would be derivation of a Landau-Lifshitz-like equation for particle motion self-consistent with the particle's radiation field.

The authors hope that this work will assist in exploring the nature of radiation-reaction
effects on firm experimental ground, and support the search for 
a fundamental theory of electromagnetism applicable in the high acceleration regime.
In that way 
the investigations presented here also reveal the possibility of reaching for
fundamental physics horizons. When the applied force approaches unity
in natural units $[m^2]$, the current physics framework has not
been tested and we are at the limit of our understanding.

\acknowledgments
Y.H., L.L. and J.R. thank Herrn Dietrich Habs, formerly the Director of the Cluster of Excellence
in Laser Physics --  Munich-Center for Advanced Photonics (MAP) for his enthusiastic
Bavarian hospitality at LMU-Garching where this research was initiated.
This work was supported by the DFG Cluster of Excellence MAP (Munich Centre of Advanced Photonics), 
and by a grant from  the U.S. Department of Energy  DE-FG02-04ER41318.

\begin{appendix} \label{AppendixSection}
\section{Supplemental formulas for the examples}\label{4vel-formulae}

a) {\bf Linearly polarized plane wave}

Equation\:(\ref{uLandauLifshitz}) for the linearly polarized plane wave in Eq.\,(\ref{LPPlaneWave}) gives
\begin{widetext}
\begin{eqnarray} \notag
\gamma &=& \frac{k \cdot u}{k \cdot u_0} \left\{1+\frac{1}{2} a_0^2\sin^2 \xi  
+(k \cdot u_0) \tau_0\left[a_0^2 \left(\!\frac{\xi}{2}-\sin\xi+\frac{3}{4}\sin 2\xi\!\right)+a_0^4 \left(\!\frac{\xi}{2}\sin^2 \xi - \frac{1}{3}\sin\xi-\frac{1}{6}\sin\xi \cos^3 \xi + \frac{1}{4} \sin 2\xi\!\right)\right]\right. \\ \notag
&& \left.+((k \cdot u_0) \tau_0)^2 \left[\frac{1}{2} a_0^2 (1-\cos\xi)^2+\frac{1}{72} a_0^4 \Big(9 (\xi +\sin \xi \cos \xi )^2-12 (\cos \xi-1)\left(-3 \xi \sin \xi+\cos ^3\xi-3 \cos \xi +2\right)\Big) \right.\right. \\ \notag
&& \left.\left.+\frac{1}{72} a_0^6 \left(-3 \xi  \sin \xi+\cos ^3\xi-3 \cos \xi+2\right)^2\right]\right\} \\ \notag
u^x &=& \frac{k \cdot u}{k \cdot u_0} \left\{a_0 \sin\xi + (k \cdot u_0) \tau_0 \left[a_0 (\cos\xi - 1)+a_0^3\frac{3\xi\sin\xi+3\cos\xi-\cos^3 \xi-2}{6}\right]\right\} \\ \notag
u^y &=& 0 \\ \notag
u^z &=& \frac{k \cdot u}{k \cdot u_0} \left\{\frac{1}{2} a_0^2\sin^2 \xi+(k \cdot u_0) \tau_0\left[a_0^2 \left(\!\frac{\xi}{2}-\sin\xi+\frac{3}{4}\sin 2\xi\!\right)+a_0^4 \left(\!\frac{\xi}{2}\sin^2 \xi - \frac{1}{3}\sin\xi-\frac{1}{6}\sin\xi \cos^3 \xi + \frac{1}{4} \sin 2\xi\!\right)\right]\right. \\ \notag
&& \left.+((k \cdot u_0) \tau_0)^2 \bigg[\frac{1}{2} a_0^2 (1-\cos\xi)^2+\frac{1}{72} a_0^4 \Big(9 (\xi +\sin \xi \cos \xi )^2-12 (\cos \xi-1)\right. 
\!\left(-3 \xi \sin \xi+\cos ^3\xi-3 \cos \xi+2\right)\Big) \\ 
&&\left.+\frac{1}{72} a_0^6 \left(-3 \xi  \sin \xi+\cos ^3\xi-3 \cos \xi+2\right)^2\bigg]\right\} .
\end{eqnarray}
b) {\bf Circularly polarized plane wave} \\
The solution in this case is
\begin{eqnarray}
\gamma &=& \frac{k \cdot u}{k \cdot u_0} \bigg\{1 + a_0^2 \left(1-\cos\xi\right)+(k \cdot u_0) \tau_0\left[a_0^2\xi + a_0^4 (\xi-\xi\cos\xi)\right] \notag \\ \notag 
&&+((k \cdot u_0) \tau_0)^2\left[a_0^2(1-\cos\xi)+a_0^4\left(\!2-2\cos\xi-\xi\sin\xi+\frac{\xi^2}{2}\!\right)+a_0^6 \left(\!1-\cos\xi-\xi\sin\xi+\frac{\xi^2}{2}\!\right)\right]\bigg\} \\ \notag
u^x &=& \frac{k \cdot u}{k \cdot u_0} \left\{-a_0(1-\cos\xi)+(k \cdot u_0) \tau_0\left[-a_0\sin\xi+a_0^3(\xi\cos\xi-\sin\xi)\right]\right\} \\ \notag
u^y &=& \frac{k \cdot u}{k \cdot u_0} \left\{a_0\sin\xi+(k \cdot u_0) \tau_0\left[a_0 (\cos\xi-1)+a_0^3 \left(\xi\sin\xi+\cos\xi-1\right) \right]\right\} \\ \notag
u^z &=& \frac{k \cdot u}{k \cdot u_0} \bigg\{a_0^2\left(1-\cos\xi\right)+(k \cdot u_0) \tau_0\left[a_0^2\xi + a_0^4 (\xi-\xi\cos\xi)\right] \\ 
&&+((k \cdot u_0) \tau_0)^2 \left[a_0^2(1-\cos\xi)+a_0^4\left(\!2-2\cos\xi-\xi\sin\xi+\frac{\xi^2}{2}\!\right)+a_0^6\left(\!1-\cos\xi-\xi\sin\xi+\frac{\xi^2}{2}\!\right)\right]\bigg\}. 
\end{eqnarray}

\section{The four-acceleration}\label{4acc-formulae}

We differentiate Eq.\,(\ref{uLandauLifshitz}) with respect to $\tau$ [remember that in Eq.\,(\ref{uLandauLifshitz}) the four-velocity is given in terms of $\xi$] and obtain the four-acceleration of the particle according to the LL equation:
\begin{eqnarray} \label{aLandauLifshitz} \notag
\dot{u}^\alpha &=& \left(\frac{k \cdot u}{k \cdot u_0}\right)^2 \left\{(k \cdot u_0) a_0 \hat{A}'^\alpha-\left[a_0 \hat{A}' \cdot u_0 + a_0^2(\hat{A}-\hat{A}_0) \cdot \hat{A}' \right] k^\alpha \right\} \\ \notag
&& +\tau_0 (k\cdot u)^2 \bigg\{\left[a_0 \hat{A}''^\alpha + a_0^2 \frac{k \cdot u}{k \cdot u_0} \hat{A}'^2 u_0^\alpha+a_0^3\frac{k \cdot u}{k \cdot u_0} \hat{A}'^2 (\hat{A}^\alpha-\hat{A}_0^\alpha)-a_0^3 \psi \hat{A}'^\alpha\right] \\ \notag
&& \left. + \bigg[-a_0 \hat{A}'' \cdot u_0 - a_0^2\hat{A}'^2-a_0^2\hat{A}' \cdot (\hat{A}'-\hat{A}'_0)-a_0^2(\hat{A}-\hat{A_0})\cdot \hat{A}'' + a_0^3 \psi \hat{A}' \cdot u_0 \right. \\ \notag
&&  - a_0^3 \frac{k\cdot u}{k \cdot u_0} \hat{A}'^2 (\hat{A}-\hat{A}_0)\cdot u_0+ a_0^4 \hat{A}' \cdot \chi + a_0^4(\hat{A}-\hat{A}_0)\cdot \hat{A}' \psi - a_0^4 \frac{k\cdot u}{k \cdot u_0} \hat{A}'^2 \frac{(\hat{A}-\hat{A}_0)^2}{2}\bigg] \frac{k^\alpha}{k \cdot u_0}\bigg\} \\ \notag
&& +\tau_0^2 (k \cdot u)^2 \bigg\{\left[a_0^3 (k \cdot u)\hat{A}'^2 (\hat{A}'^\alpha-\hat{A}'^\alpha_0)-a_0^5 (k \cdot u) \hat{A}'^2 \chi^\alpha \right] \\ \notag
&& \left. + \bigg[ -a_0^2(\hat{A}'-\hat{A}'_0)\cdot \hat{A}''-a_0^3 \frac{k \cdot u}{k \cdot u_0} \hat{A}'^2(\hat{A}'-\hat{A}'_0)\cdot u_0 + a_0^4 \hat{A}''\cdot\chi + a_0^4(\hat{A}'-\hat{A}'_0)\cdot\hat{A}' \psi +a_0^4 \psi \hat{A}'^2 - a_0^4 \frac{k \cdot u}{k\cdot u_0} \hat{A}'^2 \psi \right. \\ \notag
&&  -a_0^4 \frac{k \cdot u}{k \cdot u_0} \hat{A}'^2 (\hat{A}-\hat{A}_0)\cdot(\hat{A}'-\hat{A}'_0)+a_0^5 \frac{k \cdot u}{k \cdot u_0} \hat{A}'^2 \chi \cdot u_0 - a_0^6 \chi \cdot \hat{A}' \psi + a_0^6 \frac{k \cdot u}{k \cdot u_0} \hat{A}'^2(\hat{A}-\hat{A}_0)\cdot\chi \bigg] k^\alpha \bigg\} \\ 
&&+ \tau_0^3 (k \cdot u)^3 \hat{A}'^2 \bigg\{-a_0^4 \frac{(\hat{A}'-\hat{A}'_0)^2}{2} + a_0^6(\hat{A}'-\hat{A}'_0)\cdot\chi + a_0^6 \frac{\psi^2}{2} - a_0^8 \frac{\chi^2}{2} \bigg\} k^\alpha,
\end{eqnarray}
\end{widetext}
where we collected terms in powers of $\tau_0$ and separated them according to the direction of propagation and dependence on the intensity of the wave $a_0$ as before.

\section{Comparison with Previous Works}\label{app:solncomp}

\subsection{Analytic Comparison}
As an initial value problem, the Landau-Lifshitz Eq.\,(\ref{LLEq}) satisfies the conditions of the uniqueness and existence theorem. Therefore, one expects the evolution of the particle to be uniquely determined by the equation together with the initial conditions. This raises the question as to whether the solution given here by Eq.\,(\ref{uLandauLifshitz}) and the solution given in \cite{DiPiazza} are equivalent or even consistent.

The differences in notation between our paper and \cite{DiPiazza} are presented in Table \ref{NotationsTable} and can be used to identify our expressions with the equivalent ones given in \cite{DiPiazza}.

\begin{table}[b]\label{NotationsTable}
\caption{Notations in our paper compared with \cite{DiPiazza}}
\begin{tabular}{|c|c|}
\hline
This paper & Reference \cite{DiPiazza} \\ \hline \hline
$k^\alpha$ & $n^\alpha$ \\ \hline
$\xi = k \cdot u$ & $\phi=n \cdot u$ \\ \hline
$k \cdot u_0$ & $\rho_0$ \\ \hline
$\frac{k \cdot u}{k \cdot u_0}$ & $\frac{1}{h(\phi)}$ \\ \hline
\end{tabular}
\end{table}

In order to see equivalence between the two solutions we define the four-vector
\begin{eqnarray} \label{IVector}
\mathcal{I}^\alpha (\xi) &=& a_0 \Delta A^\alpha (\xi) + (k\cdot u_0) \tau_0 a_0 \Delta A'^\alpha (\xi) \\ \notag
&&- (k \cdot u_0) \tau_0 a_0 ^3 \chi^\alpha(\xi)
\end{eqnarray}
and the scalar quantity
\begin{equation}\label{hScalar}
h(\xi) = \frac{k \cdot u_0}{k \cdot u} = 1-(k\cdot u_0) \tau_0 a_0 ^2 \psi(\xi).
\end{equation}
They allow us to express the solution in Eq.\,(\ref{uLandauLifshitz}) as
\begin{eqnarray} \label{uLandauLifshitzAlt}
u^\alpha&=&\frac{k\cdot u}{k\cdot u_0}\left\{u_0 ^\alpha + \frac{1}{2 k\cdot u_0}\left[h^2-1\right]k^\alpha\right. \\ \notag
&&\left.-\frac{1}{k\cdot u_0}\left[k^\alpha \mathcal{I}^\beta - k^\beta \mathcal{I}^\alpha\right] u_{0,\beta}-\frac{1}{2 k\cdot u_0} \mathcal{I}^2 k^\alpha \right\},
\end{eqnarray}
where $\mathcal{I}^\alpha$ and $h$ are functions of the phase $\xi$. Using Table \ref{NotationsTable}, this can be compared with the solution given in Eq. (11) of \cite{DiPiazza}:
\begin{eqnarray} \label{uLandauLifshitzDiPiazza}
u^\alpha&=&\frac{1}{h} \left\{u_0 ^\alpha + \frac{1}{2 \rho_0}\left[h^2-1\right]n^\alpha\right. \\ \notag
&&\left.-\frac{1}{\rho_0}\left[\mathcal{I}_1 \frac{e f_1 ^{\alpha\beta}}{m} + \mathcal{I}_2 \frac{ef_2 ^{\alpha\beta}}{m}\right] u_{0,\beta}\right. \\ \notag
&& \left. -\frac{1}{2 \rho_0} \left[\xi_1^2 \mathcal{I}_1 ^2 + \xi_2^2 \mathcal{I}_2^2\right]n^\alpha \right\},
\end{eqnarray}
where
\beqn \label{IVectorDiPiazza}
\mathcal{I} (\phi) = \int _ {\phi_0} ^\phi d\varphi \left[h(\varphi) \psi_j ' (\varphi) + \frac{2}{3} \alpha \frac{\rho_0}{m} \psi_j ''(\varphi)\right]
\eeqn
and 
\beqn
f_j ^{\alpha\beta} = n^\alpha a_j ^\beta - n^\beta a_j ^\alpha.
\eeqn
Here $a_j ^\alpha$ are two four-vectors that can be associated with the polarization of the wave, and $n^\alpha$ is the wave four-vector.

To show the equivalence of Eq. (\ref{uLandauLifshitzAlt}) and Eq. (\ref{uLandauLifshitzDiPiazza}), we rewrite four-vector (\ref{IVector}) using Eq. (\ref{chiStructure}) as
\begin{eqnarray} \label{IVectorRewrite}
\mathcal{I}^\alpha (\xi) &=& a_0 \Delta \hat{A}^\alpha (\xi) - (k \cdot u_0) \tau_0 a_0 ^3 \chi^\alpha(\xi)\\ \notag
&& + (k\cdot u_0) \tau_0 a_0 \Delta \hat{A}'^\alpha (\xi)  \\ \notag
&=& a_0 \int _0 ^\xi \hat{A}'^\alpha (y)\left[1 - (k \cdot u_0) \tau_0 a_0 ^2 \psi(y)\right]dy\\ \notag
&& + (k\cdot u_0) \tau_0 a_0 \int_0 ^\xi \hat{A}''^\alpha (y)dy  \\ \notag
&=& a_0 \int _0 ^\xi \left[\hat{A}'^\alpha (y) h(y) + (k \cdot u_0) \tau_0 \hat{A}''(y)\right] dy.
\end{eqnarray}
This reveals that $\mathcal{I}^\alpha$ in our notation is nothing other than $\mathcal{I}_1 a_1 ^\alpha + \mathcal{I}_2 a_2 ^\alpha$ in the notations of \cite{DiPiazza}.

Therefore although the notations in this paper and paper \cite{DiPiazza} differ, Table \ref{NotationsTable} together with the result of Eq. (\ref{IVectorRewrite}) show that the two solutions in Eq. (\ref{uLandauLifshitzAlt}) and Eq. (\ref{uLandauLifshitzDiPiazza}) are identical and both solve the Landau-Lifshitz Eq.\,(\ref{LLEq}).

\subsection{Numerical Comparison}
In \cite{DiPiazza2} the influence of radiation reaction according to the LL equation was studied numerically for a head-on collision of an electron with a strong laser pulse. The electron was initially traveling in the negative $y$ direction with initial energy of $40\,{\rm MeV}$ and the laser wave propagating in the positive $y$ direction with frequency $\omega =1.55\,{\rm eV}$ and an intensity $I_0 = 5\times 10^{22}\,{\rm W/cm^2}$.  The laser wave was linearly polarized in the $z$ direction and subject to a sine-squared envelope with duration of $27\,{\rm fs}$, resulting in a motion in the $y-z$ plane given by Fig.~1 of \cite{DiPiazza2}.

\begin{figure}[b]
\begin{center}
\includegraphics[scale=.25]{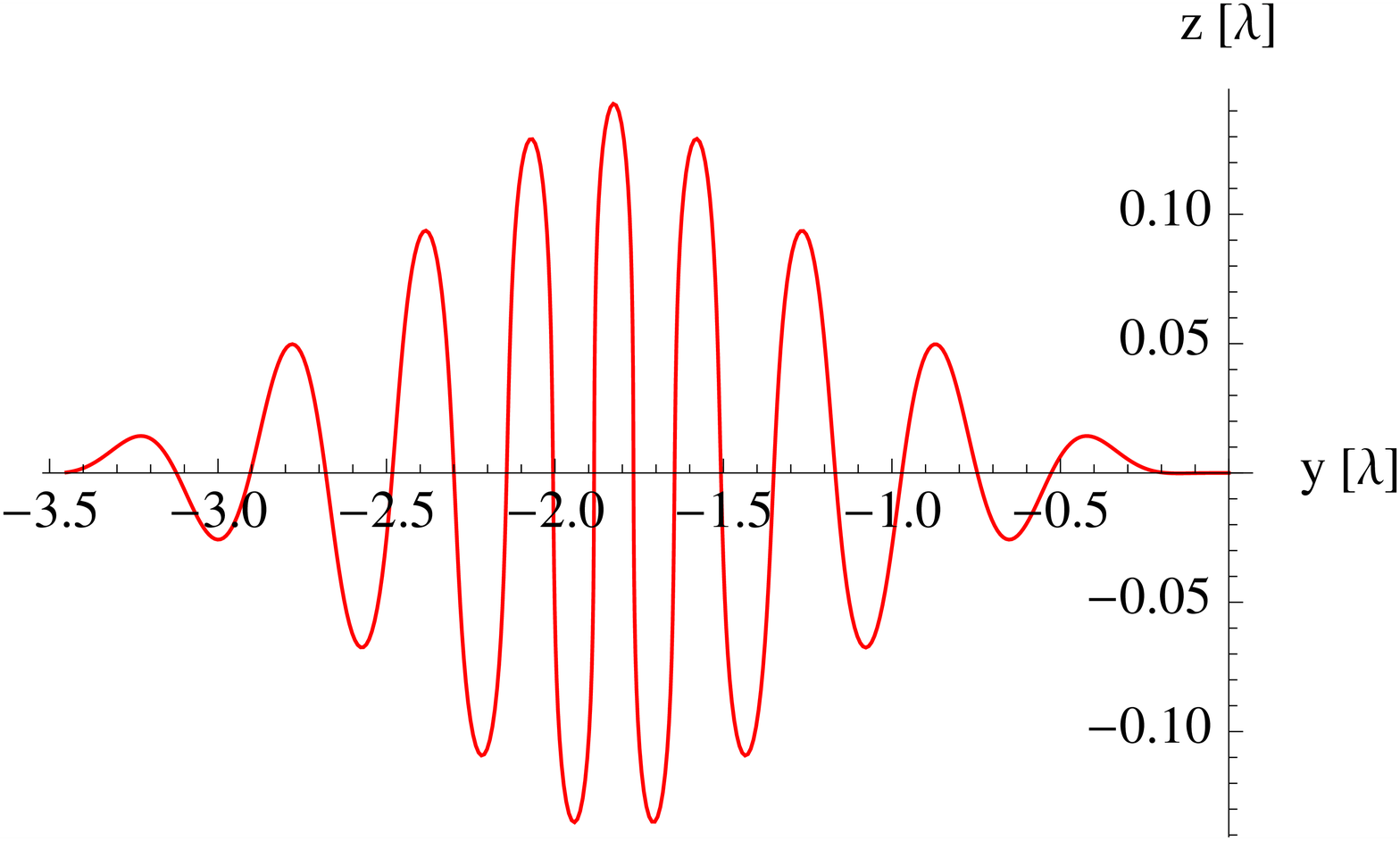}
\includegraphics[scale=.25]{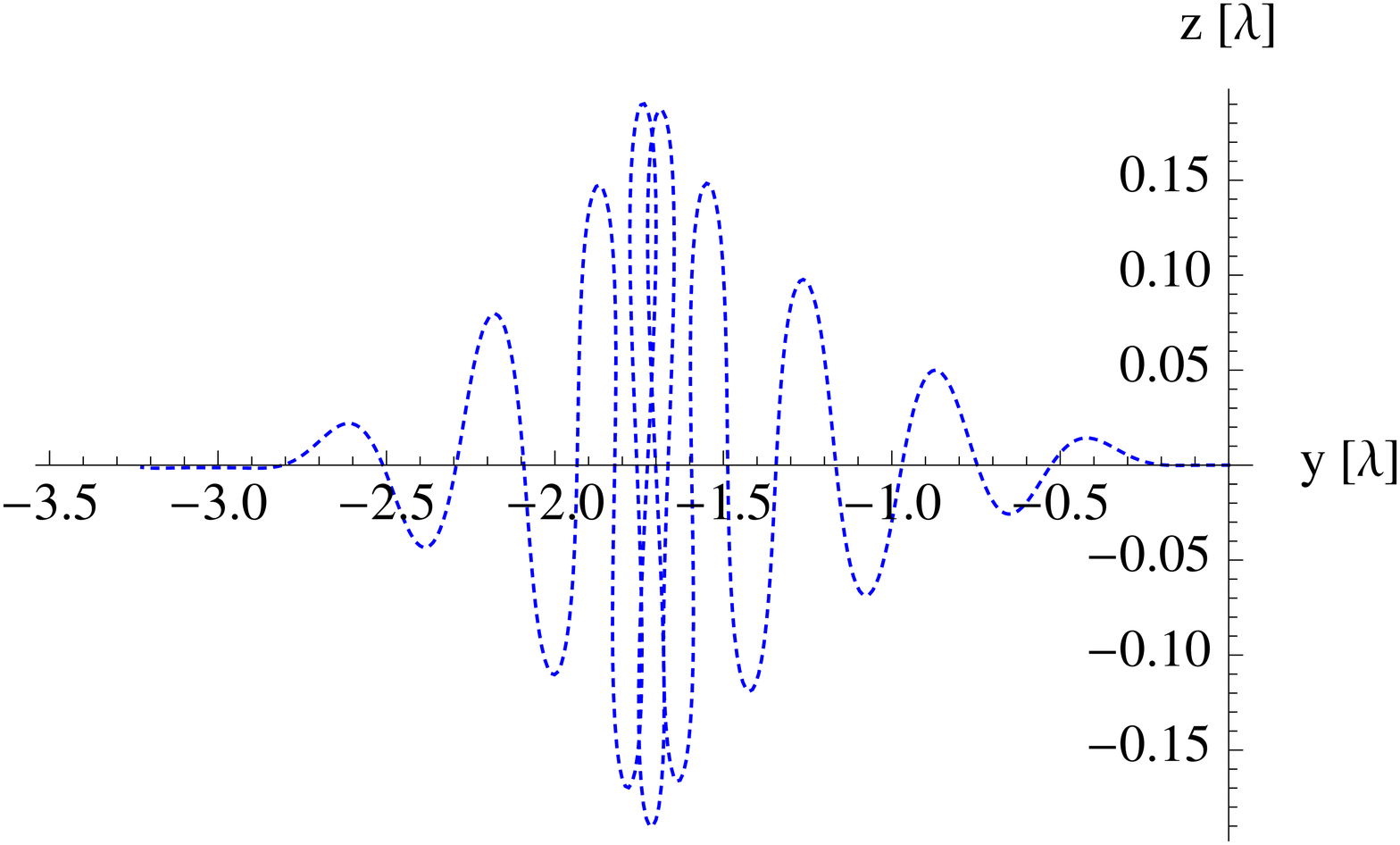}
\caption{\label{DiPiazzaFigureComparison} The analytical results for $a_0=152.925$ and $\gamma_0=78.28$ for LF (top panel) and LL (bottom panel) solutions. Compare to the numerical result in Fig.~2 of \cite{DiPiazza2} which demonstrates similar qualitative features.}
\end{center}
\end{figure}

In our notation, the above parameters translate to $\omega = 2.35\,{\rm fs}^{-1}$, $a_0 = 152.925$ and $\gamma_0=78.28$. The envelope can be modeled into the analytic solution (\ref{uLandauLifshitz}) by defining [see Eq. (\ref{WavePotential})]
\beqn
f(\xi)=-\int_0 ^\xi \cos(y) \sin^2 (y/20) dy,
\eeqn
where the cosine in the integral gives the usual linear polarization, while the sine squared term gives the required envelope shape (the factor $20$ was found by comparing it to the image in \cite{DiPiazza2}, as the exact value was not stated). Figure \ref{DiPiazzaFigureComparison} shows the corresponding analytical result obtained from Eq. (\ref{uLandauLifshitz}), and shows the same qualitative features as \cite{DiPiazza2}.

The trajectories in \cite{DiPiazza2} and Fig.~\ref{DiPiazzaFigureComparison} cannot match perfectly, as \cite{DiPiazza2} in principle considers a focused laser beam, which would have a nonplanar wave front that is not implemented in our analytic solution.  Notably, we do not find any transverse drift comparable to Fig.~1 of \cite{DiPiazza2}.  Nevertheless, the solution provided here is versatile enough to be used to provide the same features that one would expect from a focused beam.  This capability is a result of the containment apparent in Fig.\,\ref{DiPiazzaFigureComparison} of the transverse oscillation to $< 0.2\lambda$ from the beam axis, a distance over which the envelope of a focused laser pulse does not appreciably vary.

\end{appendix}

\end{document}